\documentclass[fleqn,10pt]{wlscirep_noabs}
\usepackage{amsfonts}

\title{Deterministic Generation of a Cluster State of Entangled Photons}

\author[1,+]{I. Schwartz}
\author[1,+]{D. Cogan}
\author[1,2]{E. R. Schmidgall}
\author[1]{Y. Don}
\author[1]{L. Gantz}
\author[1]{O. Kenneth}
\author[1]{N. H. Lindner}
\author[1,*]{D. Gershoni}

\affil[1]{The Physics Department and the Solid State Institute, Technion--Israel Institute of Technology, 32000 Haifa, Israel}
\affil[2]{Department of Physics, University of Washington, Seattle WA 98195, USA}

\affil[+]{These authors contributed equally to this work.}

\affil[*]{dg@physics.technion.ac.il}

\date{}

\newcommand{\ket}[1]{\left| #1 \right\rangle}             
\newcommand{\bra}[1]{\left\langle #1 \right|}             
\newcommand{\ketbra}[2]{\ket{#1}\bra{#2}}                 

\DeclareMathOperator{\Tr}{Tr}                             
\usepackage{textcomp}                                     

\usepackage{pdfpages}

\newenvironment{newabstract}{\begin{quote} \bfseries}{\end{quote}}

\begin{document}

\flushbottom

\maketitle

\thispagestyle{empty}

\begin{newabstract}
We use semiconductor quantum dots, “artificial atoms,” to implement a scheme for deterministic generation of long strings of entangled photons in a cluster state, an important resource for quantum information processing. We demonstrate a prototype device which produces strings of a few hundred photons in which the entanglement persists over 5 sequential photons. The implementation follows a proposal by Lindner and Rudolph (Phys. Rev. Lett. 2009) which suggested periodic timed excitation of a precessing electron spin as a mechanism for entangling the electron spin with the polarization of the sequentially emitted photons. In our realization, the entangling qubit is a quantum dot confined dark exciton. By performing full quantum process tomography, we obtain the process map which fully characterizes the evolution of the system, containing the dark exciton and n photons after n applications of the periodic excitations. Our implementation may greatly reduce the resources needed for quantum information processing.
\end{newabstract}

The concept of entanglement is a fundamental property of quantum mechanics~\cite{aspect1982} and an essential ingredient of proposals in the emerging technologies of quantum information processing~\cite{ladd2010, kimble2008}, including quantum communication~\cite{zukowski1993, briegel1998}
and computation~\cite{shor1994, grover1996}.
In many cases, these applications require multipartite entanglement, encompassing a large number of quantum bits (qubits). Such entanglement is often very fragile and can be adversely affected, or even completely vanish, when one of the qubits interacts with its environment or is lost from the system. A special class of quantum states exhibits a persistency of their multipartite entanglement~\cite{briegel2001}. The entanglement for this class of states is robust to adverse effects on a subset of the qubits. A prominent example of such a quantum state is a cluster state~\cite{briegel2001}---a string of mutually entangled qubits. Cluster states serve as an important resource for quantum computing, allowing its implementation solely via single qubit measurements~\cite{raussendorf2003}.
A photonic implementation of cluster states enjoys many advantages, due to the non-interacting nature of photons which suppresses decoherence effects, as well as the precise single qubit measurements provided by linear optics technology.
Generating cluster states is, however, a formidable task.
Previously, cluster states of finite size have been demonstrated in trapped ions~\cite{mandel2003,Lanyon2013},
and in continuous-variable modes of squeezed light~\cite{akira2013}.
In addition, photonic cluster states have been obtained using frequency downconversion techniques~\cite{walther2005, prevedel2007, tokunaga2008, Lu2007}.
Despite these demonstrations, the quest for obtaining a scalable, deterministic source of cluster states is still underway. 
Here, we use quantum dots, ``artificial atoms'', which are on-demand sources of both single photons~\cite{michler-nature} and entangled photon pairs~\cite{akopian2006, dousse2010, muller2014} to implement a scheme for deterministic generation of long strings of entangled photons in a cluster state~\cite{lindner2009}.  
We demonstrate a prototype device which produces strings of a few hundred photons in which the entanglement persists over $5$ sequential photons.
Feasible improvements of the device provide a route for both determinism and scalability.
Our implementation thereby forms a building block for future quantum information processing developments such as measurement-based quantum computation and quantum communication.

Our implementation is based on a proposal of Lindner and Rudolph~\cite{lindner2009}. In their proposal, repeated timed optical excitations of a confined electron in a single semiconductor quantum dot (QD) result in the formation of a cluster state composed of the sequentially emitted single photons. The proposal utilizes the spin of the electron as a matter spin qubit, whose state is entangled with the polarization of the emitted photon~\cite{akopian2006,gao2012,degreve2012,schaibley2013,rempe2014} resulting from the optical excitation. This excitation can, in principle, be repeated indefinitely, while the precessing electron spin acts as an ``entangler" and entangles the emitted photons to produce a one dimensional cluster state.

\begin{figure}[p]
\centering
\includegraphics[width=0.9\linewidth]{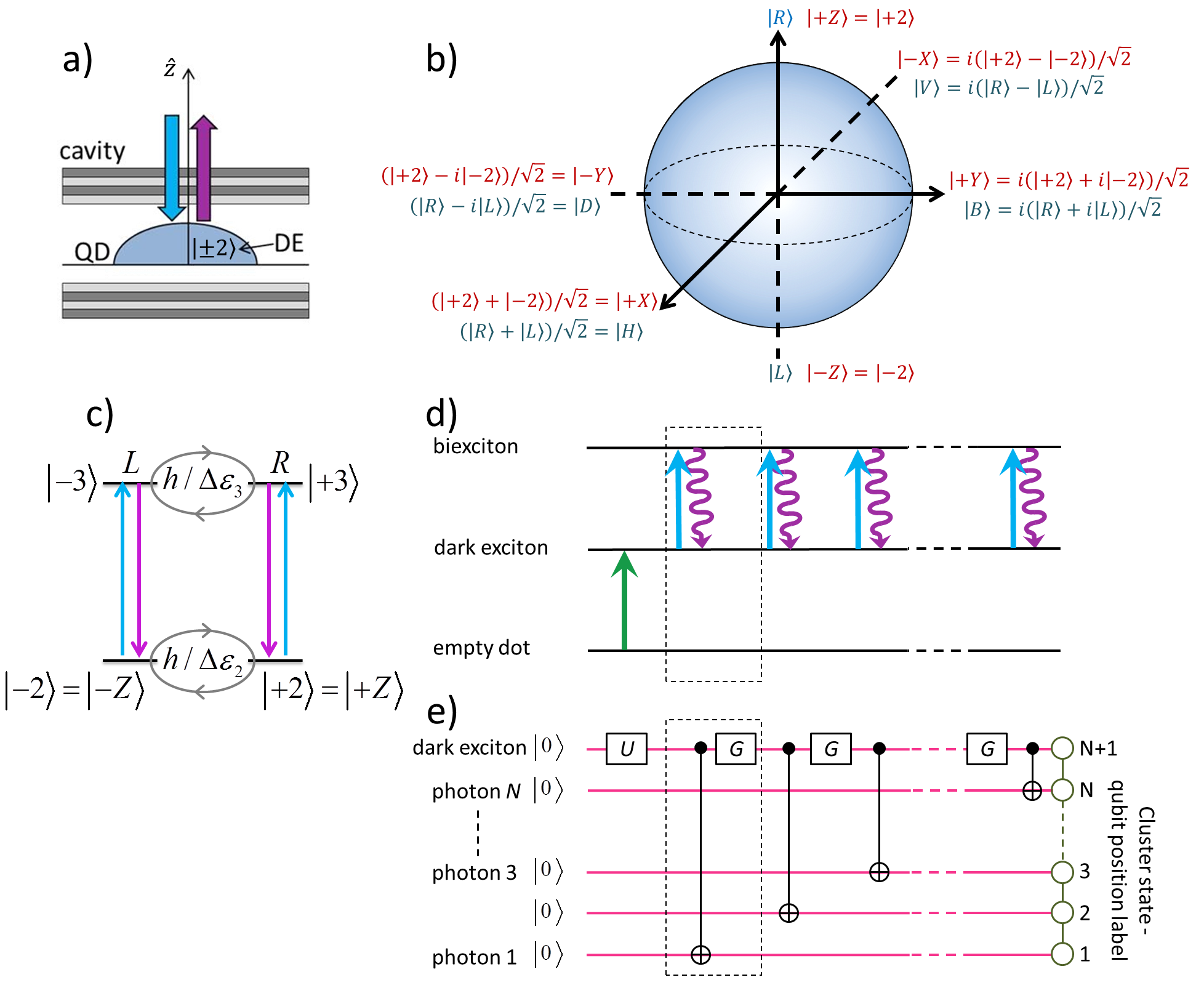}
\caption{
Schematic description of the cluster state generator.
a) The QD containing a DE in a microcavity.
The optical axis (indicated by a blue arrow for incoming laser light and a magenta arrow for outgoing emitted light) is parallel to the QD symmetry axis.
b) Bloch sphere representation of the DE spin and photon polarization qubits. The basis for the DE and photon states used throughout this work are defined in red and blue, respectively.
c) The DE states ($\ket{\pm2}=\ket{\pm Z}$), BiE states ($\ket{\pm3}$), and optical transition rules between the states. Upwards  arrows represent resonant excitation. Downwards arrows represent photon emission. The gray circular arrows represent precession of the DE and BiE states with periods of $T_\mathrm{DE}=h/\Delta \varepsilon_2 \simeq 3$~nsec and $T_\mathrm{BiE}=h/\Delta \varepsilon_3 \simeq 5$~nsec, respectively.
d) The sequence of transitions required to generate a cluster state using the DE. The green arrow represents the initialization pulse that generates a DE in spin eigenstate from an empty QD. Repeated timed excitations of the DE to the BiE (blue arrows) result in repeated emission of single photons (magenta arrows). Emission of a photon leaves a DE in the QD, which can then be re-excited. For the correct inter-pulse spacing,  the photons and the DE form an entangled cluster state.
e) The corresponding schematic circuit representation of the resulting one-dimensional string of polarization entangled photons (after Ref.~\cite{lindner2009}). Here, each horizontal line represents a qubit. The  uppermost line represent the DE and the lines below represent emitted photons ordered by their emission time.  $\ket{0}$ and $\ket{1}$ represent DE spin and photons polarization states (for the circuit diagram representation, the photons are initialized in the fiducial state $\ket{0}$).
The initialization of the DE is represented by the gate $\hat{U}$. The timed precession of the DE is represented by the single qubit gate $\hat{G}=\exp(i\frac{\pi}{4}\hat{\sigma}_x)$ gate ($\hat{\sigma}_x$ is the corresponding Pauli matrix), and excitation-emission is represented by a $\hat{C}_\mathrm{NOT}$ gate (vertical line) between the DE and the emitted photon. The area enclosed in the dashed box represents
one full unitary cycle in the ideal protocol. In the experiment, this unitary cycle is replaced by the process map (see text).
\label{fig:cluster}
}
\end{figure}

We present a practical realization of this proposal, in which the QD-confined electron is replaced by a confined dark exciton (DE)~\cite{poem2010,schwartz2014,schwartz2015}.
The DE is a semiconductor two level system, effectively forming a matter spin qubit (see the SM for more details).
The two $\pm2$ DE spin projections on the QD symmetry axis $\hat{z}$ form a basis, $\ket{\pm Z}=\ket{\pm 2}$, for the DE space [Fig.~\ref{fig:cluster}(a)~and~(b)].
The DE's energy eigenstates are $\ket{\pm X}=\left(\ket{+Z}\pm\ket{-Z}\right)/\sqrt{2}$, with an energy splitting $\Delta\epsilon_{2}$ corresponding to a precession period of $T_\mathrm{DE}=h/\Delta\epsilon_{2}\simeq3$~nsec~\cite{poem2010,schwartz2014}.
In addition to the DE, our experiment utilizes two states of a biexciton (BiE)---a bound state of two excitons---whose total spin projections on $\hat{z}$ are either $+3$ or $-3$ with a precession period of $T_\mathrm{BiE}=h/\Delta\epsilon_{3}\simeq5$~nsec. We denote these states by $\ket{\pm3}$. 
The experimental protocol relies on the optical transition rules for the transitions $\ket{+2}\leftrightarrow\ket{+3}$ and $\ket{-2}\leftrightarrow\ket{-3}$, which proceed through right $\ket{R}$ and left $\ket{L}$ hand circularly polarized photons respectively~\cite{poem2010,schwartz2014}, in direct analogy with the original proposal~\cite{lindner2009}. 
The energy level diagram describing the DE, BiE, and the optical transition rules is schematically summarized in Fig.~\ref{fig:cluster}(c).

We first present an idealized protocol for sequentially  generating the cluster state. Before the protocol begins, the DE is deterministically initialized in a spin eigenstate, $\ket{\psi_\mathrm{DE}^\mathrm{init}}=\ket{-X}$, by a short $\pi$-area picosecond pulse.~\cite{schwartz2015} A $\pi$-area pulse transfers the entire population from one quantum state to another.
The protocol, which begins immediately after the initialization, consists of repeated applications of a cycle. The cycle contains three elements: 1) a converting laser $\pi$-pulse, resonantly tuned to the DE-BiE optical transition. The pulse is rectilinearly horizontally $\ket{H}=\bigl(\ket{R}+\ket{L}\bigr)\big/\sqrt{2}$ polarized, and coherently converts the DE population to a BiE population, 2) subsequent radiative recombination of this BiE, resulting in a DE in the QD and emission of a photon,
and 3) timed free precession of the DE spin.
This cycle can be applied multiple times to generate an entangled multiphoton cluster state. Fig.~\ref{fig:cluster}(d)~and~(e) illustrate the above procedure and its equivalent circuit diagram. One full cycle of the protocol is indicated in the circuit diagram  by a dashed rectangle.

The state generated by the above protocol is revealed by following the evolution of the system during the first three steps after the initialization.
In step (1), the horizontally polarized $\pi$-area pulse ``converts" the DE coherent state into a coherent BiE state: $\ket{\psi}_\mathrm{BiE}=\bigl(\ket{+3}-\ket{-3}\bigr)\big/\sqrt{2}$.
In step (2), radiative recombination of this BiE results in an
entangled state of the emitted photon polarization and the DE spin~\cite{schwartz2014},
$\ket{\psi}_\mathrm{DE-1} = \bigl(\ket{+Z}\ket{R_1}-\ket{-Z}\ket{L_1}\bigr) \big/ \sqrt{2}$,
where $\ket{R_1}$ $(\ket{L_1})$ is the first photon right (left) hand circular polarization state.
The excitation and subsequent photon emission, are represented in Fig.~\ref{fig:cluster}(e) by a $\hat{C}_\mathrm{NOT}$ gate between the DE and the emitted photon.
In step (3), the DE precesses for $3/4$ of a precession period.
The precession is represented by the single qubit gate $\hat{G}$ (see SM) in Fig.~\ref{fig:cluster}(e). The sequential application of the $\hat{C}_\mathrm{NOT}$ and $\hat{G}$ gates forms one full cycle in our protocol. In the beginning of the next cycle, the DE-photon state is given by
\begin{equation}
	\ket{\psi}_\mathrm{DE-1} = \bigl[ \bigl(\ket{+Z}+i\ket{-Z}\bigr)\ket{R_1}-\bigl(\ket{-Z}+i\ket{+Z}\bigr)\ket{L_1} \bigr] \big/ 2.
\label{eq:DE-P1}
\end{equation}
The above cycle is now repeated: re-excitation~\cite{poem2010,schwartz2014} to the BiE state,  recombination of the second BiE and timed precession associated with the $\hat{G}$ gate. This results in a second photon, whose polarization state is entangled with that of the first photon and the spin of the remaining DE, yielding the tri-partite state
\begin{equation}
\begin{aligned}
	\ket{\psi}_\mathrm{DE-2-1} =
	\Bigl( &
		\ket{+Z} \Bigl[  \bigl(\ket{R_2}-\ket{L_2}\bigr)\ket{R_1} -i \bigl(\ket{R_2}+\ket{L_2}\bigr)\ket{L_1} \Bigr] \\
	+ & \ket{-Z} \Bigl[ i\bigl(\ket{R_2}+\ket{L_2}\bigr)\ket{R_1} +  \bigl(\ket{R_2}-\ket{L_2}\bigr)\ket{L_1} \Bigr] \Bigr) \Big/ 2\sqrt{2}.
\end{aligned}
\label{eq:DE-P2-P1}
\end{equation}
Repetition of the re-excitation--emission and subsequent precession cycle generates a one dimensional string of polarization-entangled photons in a cluster state, as shown in the equivalent circuit diagram of Fig.~\ref{fig:cluster}(e).
Here, we report on the successful realization of the above protocol in which the cycles were implemented with fidelity of 0.81 to the ideal cycle described above (see full details below).

The DE has many advantages as an entangler for sequential generation of entangled photons.
It exhibits a long lifetime $\sim 1000$~nsec and a long coherence time $T_2^*\sim 100$~nsec~\cite{schwartz2014}.
In addition, the DE spin state can be deterministically written in a coherent state using one single short optical pulse~\cite{schwartz2014,schwartz2015}, and can be reset (i.e., emptied from the QD) using fast all optical means~\cite{schmidgall2015}. Furthermore, the DE to BiE excitation resonance occurs at a higher energy than the BiE to DE main emission resonance, thereby facilitating accurate background-free single photon detection (see Methods).

In practice, however, several types of imperfections must be considered~\cite{lindner2009}. The dominant imperfection originates from the finite BiE radiative lifetime, $t_\mathrm{rad} \simeq 0.33$~nsec~\cite{schwartz2014}. 
Since the DE and BiE precess during the emission process, the purity of the polarization state of the photons is reduced~\cite{lindner2009}.
Another type of imperfection is the decoherence of the DE spin during its precession, resulting from the hyperfine interaction between the DE and nuclear spins in the semiconductor~\cite{lindner2009}.
Therefore, to ensure generation of a high quality cluster state, three important parameters should be kept small:
the ratio between the BiE radiative time $t_\mathrm{rad}$ and the DE and BiE precession times $T_\mathrm{DE}$ and $T_\mathrm{BiE}$,
 and the ratio between the DE precession time and its decoherence time $T_2^*$.
In our system, $ t_\mathrm{rad}/T_\mathrm{DE} \sim  t_\mathrm{rad}/T_\mathrm{BiE} \sim 0.1 $ and $T_\mathrm{DE}/T_2^* \sim 0.04$. Since all these parameters are much less than unity, the implemented protocol has high fidelity to the ideal one, as we now show.

The demonstration that our device generates an entangled multi-photon cluster state is done in two complementary steps.
First, we determine the non-unitary process map acting in each cycle of the protocol, which replaces the $\hat{C}_\mathrm{NOT}$ and $\hat{G}$ unitary gates of Fig.~\ref{fig:cluster}(e). The process map is a linear map from the initial DE qubit's space to the space of two qubits comprising of the DE and the newly emitted photon. It fully characterizes the evolution of the system in each cycle of the protocol, thereby completely determining the multi photon state after any given number of cycles. Then, we verify that the three qubit state, consisting of the DE and two sequentially emitted photons, generated by applying two cycles of our protocol is a genuine three qubit entangled state. We also quantify the degree of entanglement between each of the three pairs of qubits by a measure called ``localizable entanglement'' (see Methods).

To measure the process map, we perform quantum process tomography using the following procedure. We first initialize the DE in 4 different states, $\ket{\psi^\mathrm{init}_\mathrm{DE}}=$  $\ket{+X}$, $\ket{-X}$, $\ket{-Y}$, and $\ket{+Z}$. The states are defined in Fig.~\ref{fig:cluster}(b). We note that in reality the initialization is in a partially mixed state (see SM). For each DE initialization, we apply one cycle of the protocol, and perform correlation measurements between the resulting emitted photon polarization and the DE spin. In these correlation measurements, we project the emitted photon on the polarization states $\ket{H}$, $\ket{V}$, $\ket{D}$, and $\ket{R}$, while making projective measurements of the DE's spin state.
For the DE projective measurements, we apply either a right or a left hand circularly polarized $\pi$-area pulse at the end of the cycle, which due to the optical selection rules (Fig.~\ref{fig:cluster}c) deterministically excites either the $\ket{+Z}$ or the $\ket{-Z}$ DE to the BiE, respectively. Detection of an emitted photon following this excitation projects the DE spin on the states $\ket{+Z}$ or $\ket{-Z}$ at the time of the pulse. To project the DE onto the spin states $\ket{-Y}$ or $\ket{+Y}$ at the end of the cycle, we rotate the DE state by delaying the pulse by a quarter of a precession period. This method allows us to project the DE on the states: $\ket{\pm Z}$ and $\ket{\pm Y}$, but not on the $\ket{\pm X}$ states. Therefore, by the above two photon correlations, we directly measure only 48 out of the required 64 process map matrix elements.

To complete the quantum process tomography we use three photon correlation measurements. For each of the 4 initializations of the DE, we apply two cycles of the protocol. We then perform full polarization state tomography between the resulted two emitted photons, while projecting the spin of the remaining DE on $\ket{\pm Z}$ and $\ket{\mp Y}$ as discussed above.
The two and three photon correlations, which together uniquely determine all the 64 matrix elements of the process map, are fully described in the Methods section and SM.

\begin{figure}[ht]
\centering
\includegraphics[width=1\linewidth]{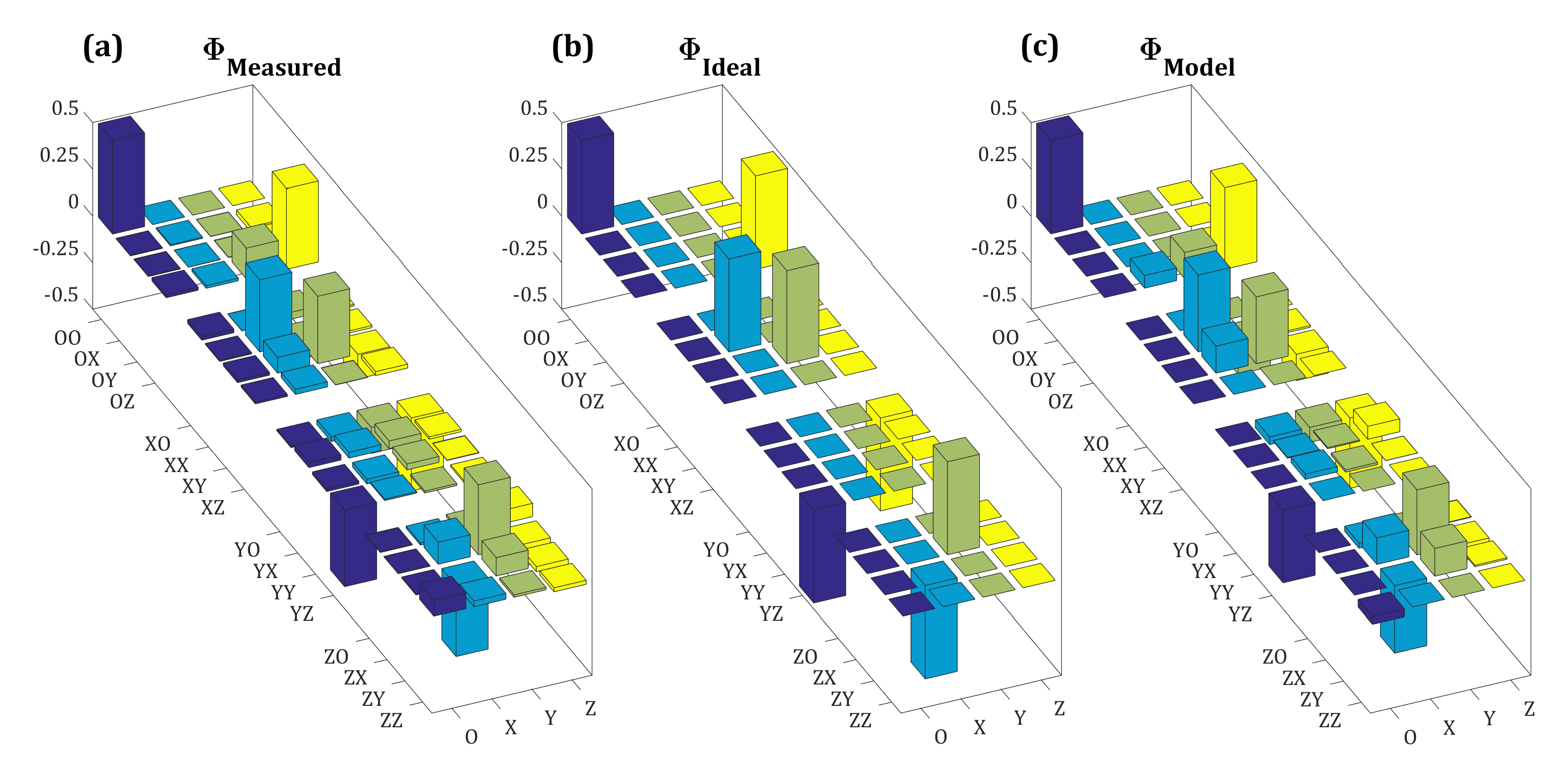}
\caption{
The process map $\Phi$ which describes the evolution of the system in each cycle of the protocol. $\Phi$ is a completely positive and trace preserving  map from the DE one qubit space to the two qubit space comprised of the DE and the emitted photon,
while acting trivially on all other photons in the state. We use the convention $\Phi\bigl(\hat{\rho}^\mathrm{(DE)}\bigr)=\sum_{\alpha\beta\gamma}\Phi^\gamma_{\alpha\beta}\rho^\mathrm{(DE)}_\gamma \hat{\sigma}_\alpha\hat{\sigma}_\beta$, where $\hat{\rho}^\mathrm{(DE)}=\sum_\gamma \rho^\mathrm{(DE)}_\gamma \hat{\sigma}_\gamma$ is the density matrix which describes the DE state. The sums are taken over $\alpha,\beta,\gamma=0,x,y,z$, where
$\hat{\sigma}_0$ is the identity matrix and $\hat{\sigma}_{x,y,z}$ are the corresponding Pauli matrices.
The 64 real parameters $\Phi^\gamma_{\alpha\beta}$, thus fully specify $\Phi$.\\
(a) The process map as \emph{measured} by quantum process tomography. The values of $\Phi^\gamma_{\alpha\beta}$, are presented such that the rows correspond to the indices $\alpha\beta$, and the columns correspond to the index $\gamma$.
\\
(b) The unitary process map corresponding to the \emph{ideal} protocol as in Fig.~\ref{fig:cluster}(e).\\
(c) The process map as calculated from a theoretical \emph{model} with independently measured parameters (see SM).
\label{fig:process}
}
\end{figure}

In Fig.~\ref{fig:process}(a) we present the measured process map, and in Fig.~\ref{fig:process}(b) we present the process map corresponding to the ideal protocol. The fidelity (see method section for definition) between the measured and ideal process map is $0.81\pm 0.1$. Fig.~\ref{fig:process}(c) gives the process map obtained by modeling the evolution of the system, using a master equation approach with independently measured system parameters (see SM). The fidelity between the model and the measured process map is $0.92\pm 0.1$.
These fidelities indicate that our device is capable of generating photonic cluster states of high quality.

Using the measured process map, we can now describe the state $\hat{\rho}_{N+1}$, consisting of DE and $N$ photons, following successive applications of $N$ protocol cycles.
An important method to quantify the multipartite entanglement in $\hat{\rho}_{N+1}$ is to consider the maximal degree of entanglement between two chosen qubits once the rest of the qubits are measured. The resulting quantity is referred to as the localizable entanglement (LE) between the two chosen qubits~\cite{localizableentanglement}. In an ideal cluster state, the localizable entanglement is maximal: any two qubits can be projected into a maximally entangled state by measuring the rest of the qubits. For example, consider the three qubit state described by Eq.~\eqref{eq:DE-P2-P1}. By projecting the DE on the $\ket{+Z}$ state, we obtain two photons in a maximally entangled state:
\begin{equation}
\begin{aligned}
	\ket{\psi}_\mathrm{2-1}
		& = \bigl[   \bigl( \ket{R_2} - \ket{L_2} \bigr) \ket{R_1}
		          -i \bigl( \ket{L_2} + \ket{R_2} \bigr) \ket{L_1}  \bigr] \big/ 2 \\
		&  =          \bigl( \ket{V_2}\ket{R_1} + \ket{H_2}\ket{L_1} \bigr) \big/ \sqrt{2} i,
\end{aligned}
\label{eq:P2-P1}
\end{equation}
where $\ket{H_2}$ and $\ket{V_2}$ are the horizontal and vertical rectilinear polarization state of the second photon.

To compute the LE between two qubits in the state $\hat{\rho}_{N+1}$, we obtain their reduced density matrix after the rest of the qubits have been measured in an optimized basis.
The degree of entanglement between the two qubits is then evaluated by a standard measure: the negativity~\cite{peres1996} $\mathcal{N}$ of the reduced density matrix. For $\mathcal{N}\ge 0$ the qubits are entangled, and $\mathcal{N}=\frac{1}{2}$ corresponds to maximally entangled qubits. Full definitions for the negativity and LE are given in the Methods section.

\begin{figure}[!ht]
\centering
\includegraphics[width=1\textwidth]{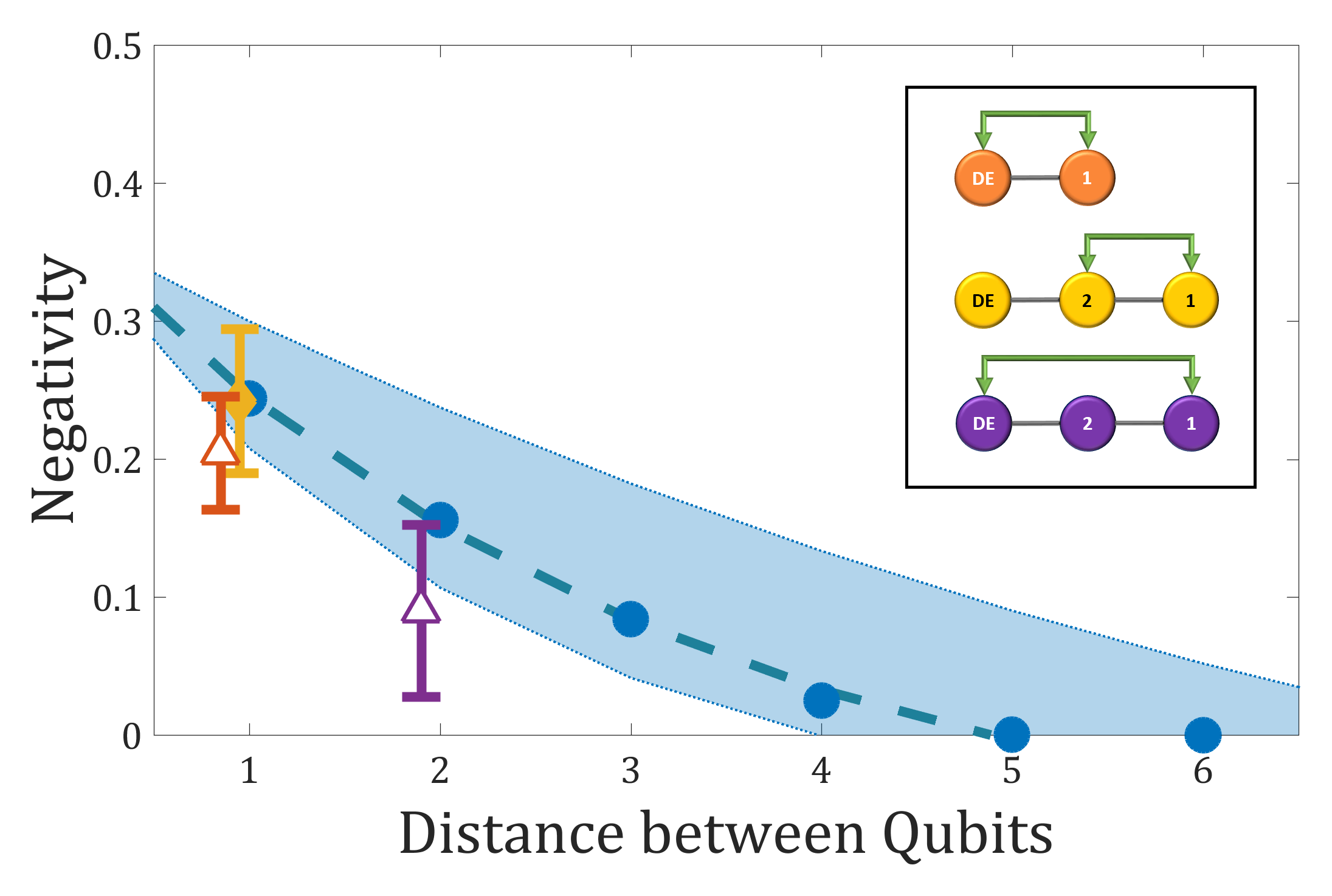}
\caption{
The localizable entanglement (LE) in the state generated by our device.  The horizontal axis indicates the distance $d$ between two qubits in the string (the qubits are ordered according to the labeling convention in Fig~\ref{fig:cluster}(e)), and the vertical axis corresponds to the negativity of the reduced density matrix of the two qubits. The blue circles correspond to $\mathcal{N}^\mathrm{LE}_{m,m+d}$ in the state $\hat{\rho}_{N+1}$ obtained using the measured process map (see Methods section for definition of the LE). Here we use $m=1$ and $d=N$, but other values for $m$ and $d\leq N+1-m$ yield similar results.
The dashed lines  represent the best fit to an exponential decay law, $\mathcal{N}^\mathrm{LE}_{m=1,m+d} = \mathcal{N}_0 \exp\left(-d/\xi_\mathrm{LE}\right)$. The blue shaded area represents one standard deviation uncertainty in the measurements determining the process map. The orange, yellow and purple data points represent the directly measured LE in two and three qubit strings, where the color coded inset describes for each data point the pair of qubits between which the LE is measured. The yellow and purple data points correspond to a lower bound on the negativity. To obtain the yellow (purple) data points, we project the DE (2nd photon) on the state $\ket{Z}$ ($e^{i(0.1)\hat{\sigma}_z}\ket{V}$).
\label{fig:localizent}
}
\end{figure}

In Fig.~\ref{fig:localizent}, we plot using blue circles the LE in the state $\hat{\rho}_{N+1}$, obtained from the measured process map, as a function of the distance between two qubits in the string. 
As expected, the LE in the one dimensional state $\hat{\rho}_{N+1}$ decays exponentially with the distance between the chosen 2 qubits~\cite{localizableentanglement}. 
The LE, which characterizes the robustness of the multipartite entanglement in the state produced by our device, is shown to persist up to 5 qubits.

Next, we used the measured correlations to directly obtain the LE between the different qubits.
First, we consider the entanglement between the DE and the emitted photon after one application of the cycle.
Using the first set of DE-photon correlation measurements we obtain a lower bound of $\mathcal{N}\geq 0.20\pm0.04$ for the negativity of their density matrix. The obtained lower bound is marked as an orange data point in Fig.~\ref{fig:localizent}. Second, we consider the three qubit state obtained after two applications of the protocol (comprising of the DE and the two emitted photons). We measure the  density matrix of the two emitted photons, while projecting the DE on the state $\ket{+Z}$.  The negativity of this measured density matrix, $\mathcal{N}=0.24\pm0.05$, is marked by the yellow data point in Fig.~\ref{fig:localizent}. The measured density matrix has fidelity of $0.81\pm0.1$ with the maximally entangled state expected from the ideal case as given by Eq.~\eqref{eq:P2-P1}.
Third, a lower bound on the LE between the DE and the first emitted photon, when the 2nd emitted photon is projected on the optimized state $e^{i(0.1)\hat{\sigma}_z}\ket{V}$, is extracted from our three photon correlations. The value of this lower bound, $\mathcal{N} \geq 0.09\pm0.06$, is marked in purple in Fig.~\ref{fig:localizent}. Note the good agreement between the LE obtained directly and the one obtained using the process map. The full set of measurements leading to the above values of $\mathcal{N}$ is given in the SM.

Finally, we use the measured DE-photon-photon correlations to directly verify that the 3 qubit state generated by our device exhibits genuine 3 qubit  entanglement~\cite{rempe2014}. Since we did not project the DE spin on the $\ket{\pm X}$ states, the density matrix of the 3 qubits cannot be fully reconstructed~\cite{gao2012,degreve2012,schaibley2013}. However, our measurements are sufficient to obtain bounds for the fidelity between the 3 qubit state $\hat{\rho}_3$ produced by our device and the 3 qubit pure cluster state $\ket{\psi}$ of Eq.~\eqref{eq:DE-P2-P1}, expected from the ideal process map. The value $\mathcal{F}=\bra{\psi} \hat{\rho}_3 \ket{\psi}$ that we obtain is $0.54 \le \mathcal{F} \le 0.69$.
The threshold for genuine 3 qubit entanglement~\cite{TothGuhne2005} is $\mathcal{F} > 1/2$ .
The experimentally measured lower bound on $\mathcal{F}$ is larger by more than one standard deviation of the experimental
uncertainty from this threshold. For full details on the calculation of $\mathcal{F}$, see the SM.

Both the quality of the state produced by our device and the correlation measurements can be significantly improved in future experiments.
Our demonstration and analysis are based on high repetition rate ($76$~MHz) deterministic writing of the entangler and on time tagged three photon correlation measurements. Direct measurements of higher multi-photon correlations are very challenging in the current experimental setup, due to its limited light harvesting efficiency, where only about $1$ in $700$ photons are detected (see SM). The limitation is mainly due to the low quantum efficiency of single photon silicon avalanche photodetectors ($< 20 \% $) and the low light collection efficiency from the planar microcavity ($< 20 \%$). Both can, in principle, be significantly improved (see~\cite{marsili2013} and~\cite{dousse2010}, respectively). In addition, the current length scale of the LE is mainly limited by the radiative lifetime of the BiE. This lifetime can be significantly shortened by designing an optical cavity to increase the Purcell factor of the device~\cite{dousse2010}, for this particular transition, only.

In conclusion, we provide an experimental demonstration of a prototype device for generating on demand, one-dimensional photonic cluster states~\cite{lindner2009}. The device is based on a semiconductor QD and utilizes the spin of the DE as an entangler. 
Our current prototype device can produce strings of a few hundred photons in which the localizable entanglement persists over $5$ sequential photons.
Further feasible optimizations of our device can enable faster and longer on-demand generation of higher fidelity cluster states. Additionally, by using two coupled QDs, it is possible to generate a two-dimensional cluster state in a similar manner~\cite{economou2010}. A two dimensional cluster state carries a promise for robust implementation of measurement-based quantum computation~\cite{raussendorf2007}.
These may propel substantial technological advances, possibly bringing practical, widespread implementations of quantum information processing closer.


\clearpage

\section*{Methods}
\subsection*{Experimental Setup}

\begin{figure}[p]
\centering{}
\includegraphics[width=1\textwidth]{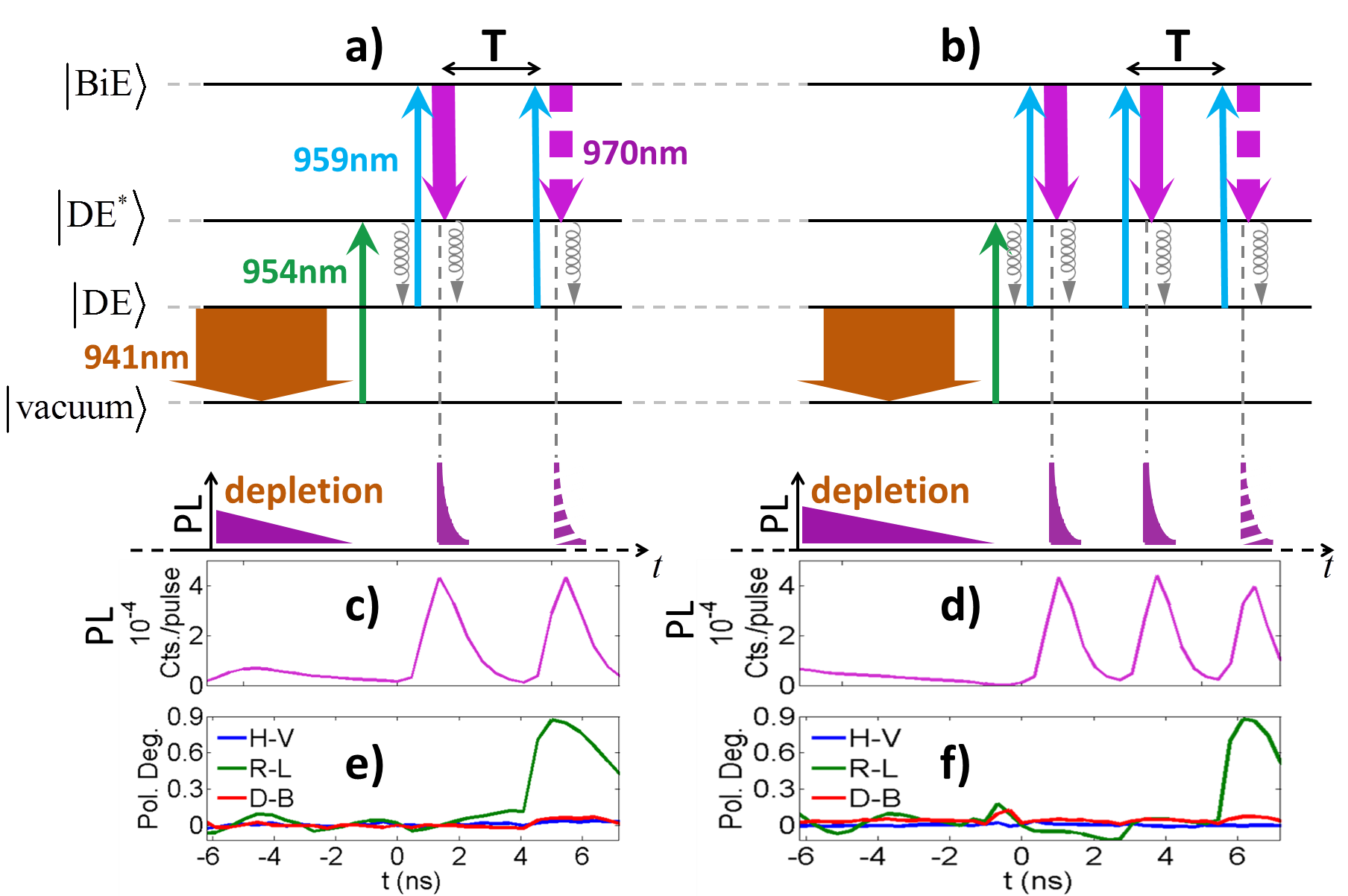}
\caption{
a) [b)] Energy levels, optical transitions, and the pulse sequence for measuring the process map by applying the protocol once [twice] using two [three] photon correlations.
The first pulse, shown as upward solid green arrow, is used to initialize the DE in one of four spin states.
The downward curly arrow represents non-radiative relaxation of the DE to its ground level. The second pulse, marked by an upward blue arrow, is an $\ket{H}$ polarized $\pi$-area pulse, which converts the DE state into a BiE state. The downward purple arrow represents the photon emission from the BiE to an excited DE level ($\mathrm{DE^*}$).
A non-radiative spin preserving relaxation of the $\mathrm{DE^*}$ to its ground level follows this emission.
The DE then freely precesses for $3/4$ of a precession period.
The conversion, emission, relaxation and precession complete one full cycle of the protocol.
A second linearly polarized $\pi$-area converting pulse and free DE precession is applied to the QD in b) as a second cycle. The last converting pulse, shown also as a solid upwards blue arrow, is $\ket{R}$ polarized, it is followed by photon emission, marked as downward dashed purple arrow. Detection of this photon projects the DE spin at the time of the polarized laser pulse on the $\ket{+Z}$ state.
This time, $T$, can be varied, as indicated by the horizontal arrow in a) and b), permitting thus DE spin projection on 3 out of 4 required, independent states.
The broad downward orange arrow, which precedes the sequence of pulses, represents a depletion pulse at $\sim 941$~nm. This pulse empties the QD from the DE~\cite{schmidgall2015}, and prepares it for a deterministic generation of the DE by the first pulse.
This sequence of pulses repeats itself at a rate of $76$~MHz.
Note that arrow colors match Fig.~\ref{fig:cluster}b).
\\
c) [d)] The emitted BiE photon count rate as a function of time as measured by each one of the four detectors. Correlated photon detections during the two [three] emission windows resulting from the two [three] converting pulses are used for the polarization tomography measurements.
\\
e) [f)] The polarization degree as a function for the time of PL signal depicted in  c) [d)].
Note that emission which follows linearly polarized pulsed conversion is unpolarized, while emission which follows circularly polarized pulsed excitation is circularly polarized.
\label{fig:exp}
}
\end{figure}

Fig.~\ref{fig:exp} describes the experimental procedures used to measure the process map by quantum tomography. The figure specifies the energy levels of the DE and BiE, the optical transitions between these levels, and the pulse sequences which we use in the measurements. Excitation pulses are indicated by upward straight arrows, subsequent emissions are indicated by downward straight arrows, and non-radiative relaxations by gray downward curly arrows.
The various arrows are color-coded according to their wavelengths, where green represents transition from the vacuum to the excited DE ($\mathrm{DE^*}$) at $\sim 954$~nm. Blue represents excitation from the DE to the BiE at $\sim 959$~nm, and magenta represents emission from the BiE to the excited DE at $\sim 970$~nm.
We note that there are two possible BiE to DE recombination paths. The oscillator strength for the transition that leaves an excited DE at $\sim 970$~nm is two orders of magnitude stronger than the oscillator strength for the transition that leaves a ground DE state in the QD.~\cite{schwartz2014}
Therefore, the BiE to ground DE emission at $\sim 954$~nm is negligible and
detected photons at $970$~nm only, are used in the polarization sensitive quantum tomography correlation measurements.
We note that coherent superpositions of the DE and BiE, which are not eigenstates of their respective two level system, precess in time
with precession times $T_\mathrm{DE}\approx 3$~nsec and $T_\mathrm{BiE} \approx 5$~nsec. These precession times are quite slow relative to the radiative lifetime of the BiE, $t_\mathrm{rad}=0.33 \pm0.03$~nsec, the non-radiative spin preserving phonon assisted relaxation time of the excited DE $t_\textrm{non-rad}=0.07\pm0.01$~nsec, and the duration of the exciting laser pulse of $12$~psec.

Fig.~\ref{fig:exp}a) schematically describes the experiment for measuring the process map by applying the protocol once and then using two photon polarization sensitive correlation measurements, and Fig.~\ref{fig:exp}b) describes the measurements when the protocol is applied twice and three photon correlations are used.
The first pulse in both figures is shown as an upward solid green arrow. This pulse is used for on-demand initialization of the DE in four states: $\ket{+X}$, $\ket{-X}$, $\ket{+Y}$ and $\ket{+Z}$, as described in Fig.~\ref{fig:cluster}b). Following rapid non-radiative relaxation $t_\textrm{non-rad}$, a second $\ket{H}$ polarized $\pi$-area pulse---marked by an upward blue arrow---converts the DE eigenstate into a BiE eigenstate. Downward purple arrow represents a photon emission from the BiE, and formation of an excited DE state. Following this emission, a non-radiative spin preserving relaxation of the DE to its ground level leaves an entangled state of the DE with the polarization of the emitted photon.
The DE then freely precesses for $3/4$ of a precession period. This 3 fold increase in the precession time relative to the original proposal~\cite{lindner2009}
is required due to the limited temporal resolution of the single photon detectors ($\approx 0.4$~nsec).
The excitation, relaxation, emission and free precession of the DE forms the repetitive process map of our protocol. In Fig.~\ref{fig:exp}b), a second linearly polarized $\pi$-area converting pulse is then applied for a second time, following free precession of the DE. Thus establishing two applications of the process map. In both cases, the emitted photons polarization are projected on four polarization states:
$\ket{H}$, $\ket{V}$, $\ket{D}$ and $\ket{R}$ as described in Fig.~\ref{fig:cluster}b).

In both cases, the remaining DE spin is projected on 2 spin directions, $\ket{\pm Y}$, and $\ket{\pm Z}$ only, using the following procedures: We use either $\ket{R}$, or $\ket{L}$ polarized converting pulse, shown as a solid upwards blue arrow. Detection of an emitted photon, after this excitation, marked as downward dashed purple arrow, projects the DE spin on the $\ket{+Z}$ or $\ket{-Z}$ state at the time of the laser pulse. This time, $T$, can be varied, as indicated by the horizontal double arrow in a) and b). Setting $T = \frac{3}{4} T_\mathrm{DE}$  therefore projects the DE spin on the $\ket{+Z}$ and $\ket{-Z}$, bases, while setting $T = \frac{1}{2} T_\mathrm{DE}$ project the DE spin on the $\ket{+Y}$ and $\ket{-Y}$ spin states.

The broad downward orange arrow, which precedes  the sequence of pulses represents a depletion pulse. The depletion pulse empties the QD from the DE, and prepares it for a deterministic generation by the first pulse. This sequence of pulses repeats itself at a rate of $76$~MHz.
At this high rate, the depletion efficiency is less than that reported in Ref.~\cite{schmidgall2015}, resulting in
measured fidelity of initialization of the DE to a pure state of $0.78\pm 0.04$ (see SM).
The optimal time separation for cluster state generation is found by experimental tuning the pulse to pulse separation resulting in excellent agreement with our theoretical model (see SM).

The emitted photons are detected by four different detectors. Each detector monitors the light at 4 different polarizations, independently (see Fig.~S3 in the SM). This arrangement facilitates efficient polarization tomography between the detected photons~\cite{kwiat2001}. Fig.~\ref{fig:exp}(c)~and~(d) show the time dependent single photon emission resulting from the recombination of the BiE during one repetition of the experiments. Fig.~\ref{fig:exp}(e) and (f) show the time dependent degrees of rectilinear, diagonal and circular polarizations of the signal in Fig.~\ref{fig:exp}(c)~and~(d), respectively. We note, as expected, that emission which follows linearly polarized conversion is unpolarized, while emission which follows circularly polarized conversion is circularly polarized.

\subsection*{Measures of Entanglement}
\subsubsection*{Negativity}
Consider a density matrix describing the state of two qubits
$\hat{\rho} = \sum \rho_{ m \mu ; n \nu} \ketbra{m}{n} \otimes \ketbra{\mu}{\nu}$,
where the labels $m,n$ correspond to qubit $A$, and $\mu,\nu$ belong to qubit $B$, and the sum is over the labels $m,n,\mu,\nu = 0,1$. The partial transpose of $\hat{\rho}$ is defined by
$\hat{\rho}^{T_\mathrm{P}} = \sum \rho_{m\mu ; n\nu}  \ketbra{m}{n} \otimes (\ketbra{\mu}{\nu})^T
                      = \sum \rho_{m\nu ; n\mu} \ketbra{m}{n} \otimes \ketbra{\mu}{\nu}$.
The state $\hat{\rho}$ is entangled if and only if any of the eigenvalues of $\hat{\rho}^{T_\mathrm{P}}$, is negative~\cite{peres1996}.
Denoting by $\lambda_j$ the 4 eigenvalues of $\hat{\rho}^{T_\mathrm{P}}$, the negativity~\cite{Negativity} $\mathcal{N}$ is defined as $\mathcal{N}(\hat{\rho}) = \tfrac{1}{2} \bigl( \sum_j|\lambda_j|-\lambda_j \bigr)$, i.e., $\mathcal{N}(\hat{\rho})$ is the sum of the absolute value of the negative eigenvalues of $\hat{\rho}^{T_\mathrm{P}}$. The negativity measures the degree of entanglement in $\hat{\rho}$, where $\mathcal{N}=0$ corresponds to an unentangled state, and $\mathcal{N}=\frac{1}{2}$ to a maximally entangled state.

\subsubsection*{Localizable Entanglement}
\label{sec: LE}
Given a state $\hat{\rho}_N$ of $N$ qubits, we define the  localizable entanglement~ \cite{localizableentanglement} (LE) of two qubits $n$ and $m$ as the maximal negativity $\mathcal{N}$ of their reduced density matrix  obtained after performing projective single qubit measurements on all the other $N-2$ qubits. We denote by $\mathcal{M}$ a choice for the basis for the projective measurements, and by $s$ the possible outcomes of the measurements.   For a specific $\mathcal{M}$ and $s$, the reduced density matrix between qubits $n$ and $m$ is given by
\begin{equation*}
	\hat{\rho}_{n,m}^{(\mathcal{M},s)}=\frac{1}{p_s}\Tr_{j\neq m,n}\biggl[\biggl( \prod_{j\neq m,n} \hat P_j \biggr) \hat{\rho}_N\biggr].
\end{equation*}
In the above, the basis $\mathcal{M}$ and the outcomes $s$ correspond to the sequence of projections $\prod_{j\neq m,n} \hat{P}_j $, where $\hat{P}_j=\ket{s_j}\bra{s_j}$ is a projector acting on qubit $j$.  Note that all qubits are projected except for qubits $n$ and $m$. Likewise, the partial trace is taken over all qubits except photons $m$ and $n$. Finally, $p_s$ is the probability to obtain the measurement outcome $s$, given by $p_s=\Tr\left[\prod_{j\neq m,n} \hat P_j  \hat{\rho}_N\right]$. The average localizable entanglement between the qubits $n$ and $m$, maximized over all possible measurement basis is then given by~\cite{localizableentanglement}
\begin{equation*}
	\mathcal{N}^\mathrm{LE}_{n,m}(\hat{\rho}_{N})=\max_\mathcal{M} \sum_s p_s \mathcal{N}\left(\hat{\rho}_{n,m}^{(\mathcal{M},s)}\right).
\end{equation*}

\subsubsection*{Fidelity}
\label{subsec: fidelity}
The fidelity $\mathcal{F}$ between two Hermitian positive matrices $A$ and $B$ is defined as~\cite{jozsa1994}
\begin{equation*}
	\mathcal{F}(A,B) = \Tr\Bigl[ \sqrt{\sqrt{B}\,A\sqrt{B}} \,\Bigr]^{2}
	                   \Big/ \Bigl( \Tr\left[A\right] \, \Tr\left[B\right] \Bigr).
\end{equation*}
The fidelity between two quantum states $\hat{\rho}_1$ and $\hat{\rho}_2$  is given by $\mathcal{F}\left(\hat{\rho}_1,\hat{\rho}_2\right)$. The fidelity between two process maps $\Phi_1$ and $\Phi_2$ is given by $\mathcal{F}\left(C^{\Phi_1},C^{\Phi_2}\right)$, where $C^{\Phi}$ is the Choi matrix corresponding to $\Phi$ (see SM for the definition of the Choi matrix).

\clearpage

\bibliography{DE_cluster}


\section*{Acknowledgements}
We are grateful to Pierre Petroff for the sample growth and to Terry Rudolph and Joseph Avron for useful discussions.
The support of the Israeli Science Foundation (ISF), the Technion's RBNI, and the Israeli Nanotechnology Focal Technology Area on ``Nanophotonics for Detection" are gratefully acknowledged.
The authors declare that they have no competing financial interests. Correspondence and requests for materials should be addressed to \href{mailto:dg@physics.technion.ac.il}{dg@physics.technion.ac.il}.


\newcounter{mainpage}                  
\setcounter{mainpage}{\value{page}}    
\stepcounter{mainpage}                 


\section*{Supplementary material}

# Supplementary Materials for
## Deterministic Generation of a Cluster State of Entangled Photons

I. Schwartz[1,+], D. Cogan[1,+], E. R. Schmidgall[1,2], Y. Don[1], L. Gantz[1], O. Kenneth[1], N. H. Lindner[1], and D. Gershoni[1,*]

[1]The Physics Department and the Solid State Institute, Technion–Israel Institute of Technology, 32000 Haifa, Israel
[2]Department of Physics, University of Washington, Seattle WA 98195, USA
[+]These authors contributed equally to this work.
[*]dg@physics.technion.ac.il

## 1 Sample Description

The sample used in this work was grown by molecular beam epitaxy on a (001) oriented GaAs substrate. One layer of strain-induced $In_xGa_{1-x}As$ quantum dots (QDs) was deposited in the center of a one-wavelength microcavity. The microcavity formed by two unequal stacks of alternating quarter-wavelength layers of AlAs and GaAs. The thinner stack was deposited on top of the QD layer, thus facilitating prefered light emission towards the sample surface. The height and composition of the QDs were controlled by partially covering the InAs QDs with a 3 nm layer of GaAs and subsequent growth interruption.[1] To improve photon collection efficiency, the microcavity was designed to have a cavity mode which matches the QD emission due to ground-state $e$-$h$ pair recombinations. The sample structure is schematically described in Fig. S1. During the growth of the QD layer the sample was not rotated, resulting in a gradient in the density of the formed QDs.[1] The estimated QD density in the sample areas that were measured is $10^8$ cm$^{-2}$; however the density of the QDs that emit in resonance with the microcavity mode is much lower.[2] Thus, single QDs separated by a few tens of micrometers were easily located by scanning the sample surface during confocal photoluminescence (PL) measurements.

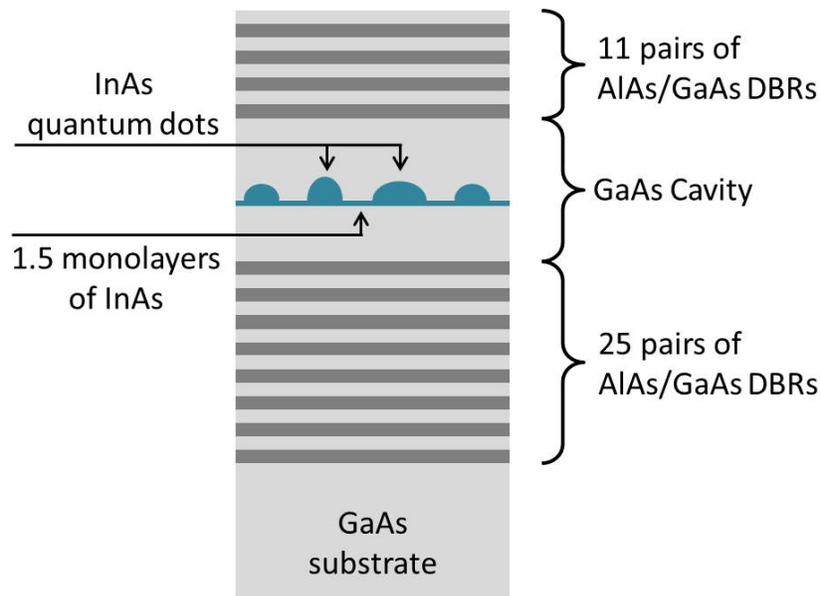

**Figure S1.** Schematic description of the studied sample. DBR indicates a distributed Bragg reflecting mirror.

## 2 The Dark Exciton Physical System

An exciton is a fundamental semiconductor excitation in which one electron from the full valence band (formed of atomic orbitals with unit angular momenta) is excited to the empty conduction band (formed of atomic orbitals with zero angular

1

momentum). The missing electron is commonly called a "hole." The hole has positive charge and opposite angular momentum and spin to that of the missing electron.[3] In QDs, due to the strain and quantum size effects, the lowest energy valence electrons (highest energy holes) have parallel spins and angular momentum.[3] Therefore while conduction electrons have a total spin projection of either $+1/2$ or $-1/2$ on the QD symmetry axis ($\hat{z}$), those of the holes are either $+3/2$ or $-3/2$.

If the electron and hole spins are antiparallel, they can recombine radiatively by photon emission (the excited electron simply "returns" to the missing electron state in the valence band). The exciton is called "bright", it lives for a short time ($\sim 0.5$ nsec), and it has integer spin projections of either $+1$ or $-1$ on the symmetry axis. If, however, the electron and hole spins are parallel, the exciton is predominantly optically inactive. It is called "dark" (DE), has either $+2$ or $-2$ integer total spin projections and it lives orders of magnitude longer ($\sim 1$ μsec[4,5]) than the BE. The exchange interaction between the electron and hole removes the degeneracy between the bright and dark excitons ($\sim 300$ μeV in this case), and also between the two spin eigenstates of the bright ($\sim 34$ μeV) and dark ($\sim 1.4$ μeV) exciton.[6] Notice that the DE eigenstates, in increasing energy order, are the antisymmetric $(|+2\rangle - |-2\rangle)/\sqrt{2}$ and symmetric $(|+2\rangle + |-2\rangle)/\sqrt{2}$ coherent superpositions of its spin states $|\pm 2\rangle = |\pm Z\rangle$[7,8] (See Fig. S2 and Fig. 1 in the main text).

The lowest energy optical excitation of the DE results in a formation of a biexciton (BiE), composed of two electron-hole pairs. The pair of electrons forms a spin singlet in their ground level while the holes form a spin-parallel triplet with one hole in the QD ground level and one in the first excited level. Thus, the BiE forms a two level system with spin states $|\pm 3\rangle$. The BiE has a relatively short lifetime ($\approx 0.33$ nsec) determined by the efficient radiative recombination of an electron with a ground level hole. This recombination results in an emitted photon at 970 nm and an excited DE (DE*) in the QD. The recombination of an electron with the excited level hole is about 50 times less efficient. This is due to the reduced overlap between the electron and hole envelope wavefunctions of different spatial symmetries. The DE* then relaxes non-radiatively to the ground DE level within $\sim 0.07$ nsec by emitting a spin-preserving phonon.[8] The DE and the BiE levels, their spin wavefunctions and the transitions between these levels are illustrated in Fig. S2.

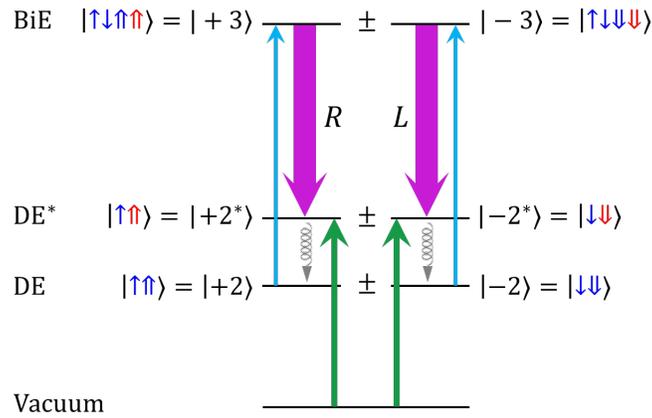

**Figure S2.** Schematic description of the DE-BiE energy levels, spin wavefunctions and optical transitions between these levels. In the spin wavefunctions $\uparrow$ ($\Downarrow$) represents spin up (down) electron (hole) with spin projection of +1/2 (-3/2) on the QD $\hat{z}$ axis. Blue (red) color in the spin wave functions represents carrier in the ground (excited) energy level. The vertical downward (upward) arrows denote resonant optical emission (excitation) transition, where $R$ ($L$) stands for right (left) hand circular polarization. The colors of the optical transitions match these of Fig. 1 and Fig. 4 in the main Letter. Thicker arrows represent stronger optical transitions. Curly arrows represent non-radiative spin-preserving hole relaxation.

## 3 Experimental Methods

The sample was placed inside a sealed metal tube cooled by a closed-cycle helium refrigerator maintaining a temperature of 4 K. A ×60 microscope objective with numerical aperture of 0.85 was placed above the sample and used to focus the light beams on the sample surface and to collect the emitted PL. Pulsed laser excitations were used. The picosecond pulses were generated by two synchronously pumped dye lasers in repetition rate of 76 MHz. The temporal width of the dye lasers' pulses was $\sim 12$ psec and their spectral width $\sim 100$ μeV. Light from a continuous wave external cavity laser, modulated by an electro-optic modulator, synchronized with the dye lasers, was used to produce the depletion pulses of a few nanosecond duration.[9] The energy of one of the picosecond long laser pulses was independently tuned to resonantly excite (write) the

DE[4] while the energy of the other one was tuned to the DE-biexciton (BiE) optical transition for excitation of the BiE (for converting and DE spin projection). The latter was divided into 3 different beams. The timing of the four picosecond long pulses was set by computer controlled motorized delay lines. The polarizations of the pulses were independently adjusted using polarized beam splitters (PBS) and three pairs of computer-controlled liquid crystal variable retarders (LCVRs). The collected PL was equally divided into 2 beams by a non-polarizing beam splitter (NPBS). Two pairs of LCVRs and a PBS were then used to analyze the polarizations of each beam. Thus, the emitted PL was divided into four beams, allowing selection of two independent polarization projections and their complementary polarizations. The polarization projected PL from each beam was spectrally analyzed by either a 1 or 0.5 meter monochromator and detected by a silicon avalanche photodetector coupled to a PicoQuant HydraHarp 400™ time-correlated single photon counter synchronized with the pulsed lasers. This way the arrival times of four emitted photons were recorded relative to the synchronized laser pulses. A special software package was developed in order to handle the data acquisition on-line. In addition, a cooled charged coupled array detector could be used for temporally integrated spectral analysis. The experimental system is schematically described in Fig. S3.

The overall detection efficiency of the system was about 1 photon in 700 (integrated over the four detectors).[10] Thus, at a repetition rate of 76 MHz, we detected about 100000, 150, and 0.2 one, two and three photon events per second, respectively. An event here is a detection of a photon (a "click" in a detector) during a 13.2 nsec single period of the 76 MHz repetition rate.

In Fig. S4, we present the time-resolved PL signal for (a) one-photon events, (b) two-photon events, and (c) three photon events for summation over all the polarization combinations required for the quantum tomography. The time resolved BiE PL emission as seen in one of the four detectors is presented in Fig. S4(a). The signal at negative times results from the depletion pulse,[9] which empties the QD from the DE as evident from the vanishing signal at $t=0$. We set $t=0$ to the time of arrival of the DE initialization pulse (green arrows in Fig. 4(a) and (b) in the main paper). The emission, which results from the sequential two converting (solid purple arrows in Fig. 4(b) in the main paper) and one projective measurement (dashed purple arrow in Fig. 4(b) in the main paper) pulses, is observed at positive times. In Fig. S4(b) and (c) temporally correlated two and three photon events are observed as dense areas in which two and three detectors, respectively, clicked during the same period. Statistics for the polarization tomography measurements are obtained from the main two and six dense regions in (b) and (c) by taking into account the temporal order by which the two and three photons were detected, respectively. The probability that two photons are emitted during one conversion pulse is measured to be 1.8%. Such low probability has no influence on the conclusions drawn in this work.

## 4 Theoretical Model

We used a Markovian master equation model to describe the evolution of the density matrix $\hat{\rho}$. The equation takes the form $\frac{d}{dt}\hat{\rho} = -\frac{i}{\hbar}[\mathcal{H},\hat{\rho}] + \mathcal{L}[\hat{\rho}]$, where $\mathcal{H}$ is the free Hamiltonian describing the system and $\mathcal{L}(\hat{\rho})$ is the Lindblad superoperator describing the radiative decay and decoherence processes.

In the following paragraphs we will describe (i) the initialization process of the DE; (ii) the excitation process from the DE to the BiE; (iii) the free evolution, decay, and decoherence of the system; and (iv) the combination of several protocol cycles required for generating a cluster state of an arbitrary length. Here, each cycle contains an excitation pulse of the DE to the BiE and the subsequent decay and temporal evolution, as depicted in Fig. S6.

### 4.1 Initialization and Reset Efficiency for the DE

We initialize the DE using a resonant picosecond-long polarized pulse to the excited dark exciton state (DE*).[4] The DE* then decays non-radiatively within 0.07 nsec, by spin preserving acoustic phonon emission to the ground DE state, while retaining its coherent state. Thus, we effectively initializing the DE state. The state in which the DE is initialized in the experiment is described in our model by: $\hat{\rho}_{\text{DE}}^{\text{init}} = p|\psi_{\text{DE}}^{\text{init}}\rangle\langle\psi_{\text{DE}}^{\text{init}}| + (1-p)\frac{1}{2}\mathbb{I}$, where $\mathbb{I}$ is the identity matrix, and $p$ is the probability of $\hat{\rho}_{\text{DE}}^{\text{init}}$ to be in the pure state $|\psi_{\text{DE}}^{\text{init}}\rangle$ dictated by the polarization of the initialization pulse.[4] For example, $|\psi_{\text{DE}}^{\text{init}}\rangle = \frac{1}{\sqrt{2}}(|+Z\rangle - |-Z\rangle) = |-X\rangle$, in the protocol for cluster state generation (Fig. 1 in the main Letter). Thus, $1-p$ describes the probability that our depletion did not result in an empty QD, and the remaining DE is in a totally mixed state. The depletion efficiency is deduced by comparing the population of the DE between the beginning and the end of the depletion pulse. We obtained $p = 0.72$. We note that the initialization probability in the current experiment is lower by about 20% than that achieved in Refs.[4,9], due to the 8 fold increase in the repetition rate reported here. The increase in the repetition rate is required for obtaining the needed statistics for the three photon correlations.

### 4.2 Excitation Pulse

We now describe the model of the population transfer from the DE to the BiE using a short excitation pulse in a single cycle. The free Hamiltonian used for describing the unitary evolution of the system is given by $\mathcal{H}_2 \oplus \mathcal{H}_3$, where $\mathcal{H}_2 = \frac{1}{2}\begin{pmatrix} 0 & \hbar\omega_2 \\ \hbar\omega_2 & 0 \end{pmatrix}$ acting on the DE states $|\pm 2\rangle \equiv |\pm Z\rangle$, and $\mathcal{H}_3 = \frac{1}{2}\begin{pmatrix} 0 & \hbar\omega_3 \\ \hbar\omega_3 & 0 \end{pmatrix}$ acting on the BiE states $|\pm 3\rangle$. Here $\hbar\omega_2 = \Delta\varepsilon_2$ and $\hbar\omega_3 = \Delta\varepsilon_3$ are

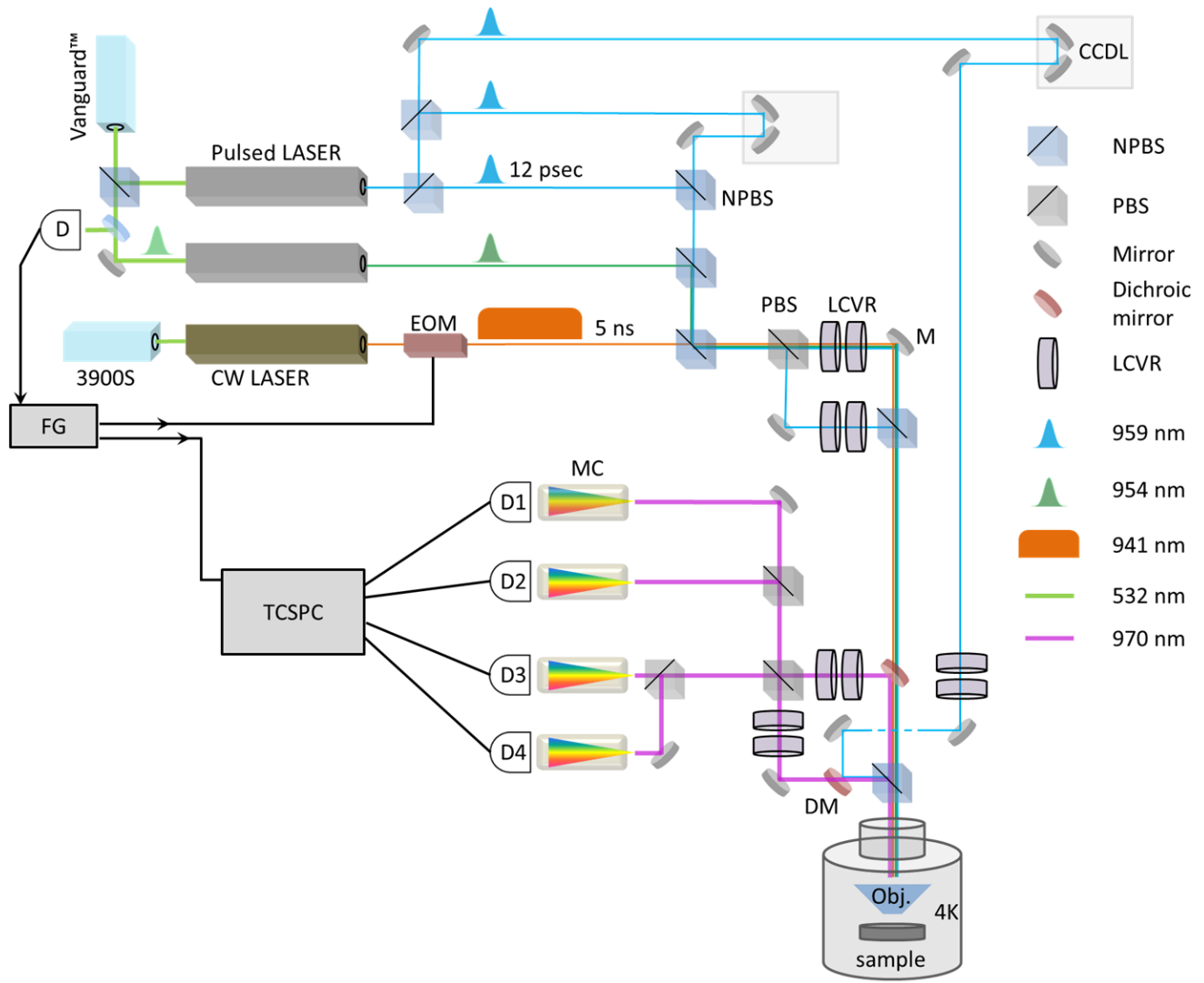

**Figure S3.** Schematic description of the experimental setup. NPBS, non-polarizing beam splitter; PBS, polarizing beam splitter; CCDL, computer controlled optical delay line; M, mirror; DM, dichroic mirror; TCSPC, Time correlated single photon counter; MC, monochromator; OBJ, microscope objective; LCVR, liquid crystal variable retarder; D$n$, detector $n$; CW, continuous wave; FG, function generator; EOM, electro-optic modulator
.

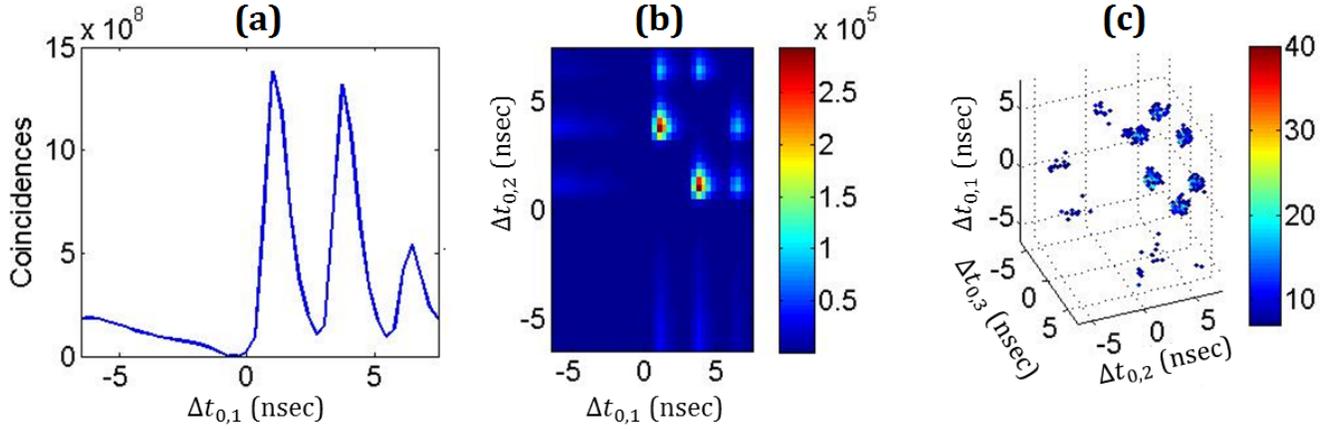

**Figure S4.** Simultaneously measured time-resolved (a) one-photon, (b) two-photon and (c) three-photon events, where the number of detected photons per time bin of 0.33 nsec. for each detector is shown as (a) the ordinate, (b) and (c) the false color scale. $\Delta t_{0,n}$ represents the time difference between the clock pulse and detection of a photon on the $n$-th detector, in units of nsec. For clarity, in (c) the threshold was set to 7 counts per three dimensional temporal bin. For each additional detection channel, the number of detected events is reduced by a factor of $\sim 1/700$.

The three peaks in (a) correspond to detections of emitted photons after each of the respective excitation pulses. The two main dense areas in (b) correspond to detected two-photon events, during one period of the 76MHz experiment. The upper-left (lower-right) dense area corresponds to events in which detector #1 (#2) detects a photon first and detector #2 (#1) detects a photon second. The four additional dense areas correspond to events in which a photon from the third pulse was detected. Similarly, the six main dense areas in (c) correspond to detection of three photon events during one period of the 76MHz experiment by three detectors.

the energy splittings of the DE and BiE eigenstates, respectively. Note that the emitted photons do not change their polarization state during the QD free evolution. During the excitation pulse the following Hamiltonian was used to describe the pulse interaction with the QD:

$$\mathcal{H}_P = \frac{1}{2} \begin{pmatrix} +\Delta & \hbar\omega_2 & 2\hbar\Omega_R(t) & 0 \\ \hbar\omega_2 & +\Delta & 0 & 2\hbar\Omega_L(t) \\ 2\hbar\Omega_R(t) & 0 & -\Delta & \hbar\omega_3 \\ 0 & 2\hbar\Omega_L(t) & \hbar\omega_3 & -\Delta \end{pmatrix}, \quad (1)$$

where $\Delta = hc/\lambda - (\varepsilon_3 - \varepsilon_2) = 0$ is the energy detuning of the resonant pulse, $\lambda$ is the excitation wavelength, and $\varepsilon_3$ and $\varepsilon_2$ are the energies of the BiE and DE, respectively; $\Omega_R(t)$ and $\Omega_L(t)$ describe the temporal shape of the right- and left-hand circular polarization components of the $\Theta = \pi$-area pulse. We assume a Gaussian pulse with width of $\hbar\Omega_W \sim 100$ µeV and define the polarization of the excitation pulse polarization as $\hat{e} = e_R \hat{R} + e_L \hat{L}$ with $e_R^2 + e_L^2 = 1$. Thus, the temporal shapes are given by: $\Omega_{R/L}(t) = \frac{1}{2}\Theta \cdot \frac{1}{\sqrt{2\pi/\Omega_W}} \exp\left(-\frac{1}{2}\Omega_W^2 t^2\right) \cdot e_{R/L}$. The system energy levels and transition rates between these levels are schematically described in Fig. S5.

### 4.3 Free Evolution and Decay

Here, we describe the decay of the BiE to the DE together with all other decoherence processes. The Lindblad superoperator takes the standard form $\mathscr{L}[\hat{\rho}] = \sum_k \hat{C}_k \hat{\rho} \hat{C}_k^\dagger - (\hat{C}_k^\dagger \hat{C}_k \hat{\rho} + \hat{\rho} \hat{C}_k^\dagger \hat{C}_k)/2$, where each operator $\hat{C}_k$ describes a different decay or decoherence process.

The decay from the BiE to the DE is described by the operators $\hat{C}_{d,R} = \sqrt{\gamma_d} |+Z, R_n\rangle \langle +3|$ and $\hat{C}_{d,L} = \sqrt{\gamma_d} |-Z, L_n\rangle \langle -3|$, corresponding to emission of a right- and left-hand circularly polarized photon, respectively; $|R_n\rangle$ and $|L_n\rangle$ are the $n$-th photon right- and left-hand circular polarization states. For simplicity, the decay rate $\gamma_d$ in our model accounts for both (i) the radiative decay from the BiE to the excited DE (determined from the measured $0.33 \pm 0.03$ nsec decay time using a pump-probe method, not shown); and (ii) the non-radiative decay of the DE* to the DE (determined from the measured $0.07 \pm 0.01$ nsec decay time using pump-probe method, not shown). We checked numerically that since the decay is spin preserving and since both decays are much shorter than the DE* and DE precession times, this simplification is justified.

The dephasing of the DE state is described by the operator $\hat{C}_2^* = \sqrt{\gamma_2^*}(|+Z\rangle\langle+Z| - |-Z\rangle\langle-Z|)$. For the rate $\gamma_2^* = 1/T_2^*$

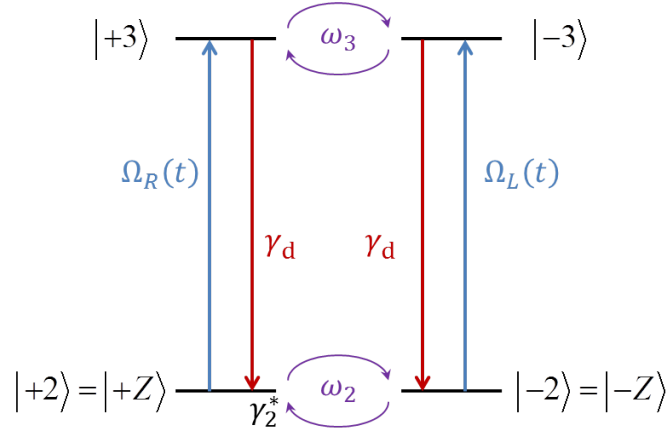

**Figure S5.** Schematic description of the energy levels and transition rates used in the model calculations. Here $|\pm J\rangle$ represents a state with total angular momentum projection $J$ on the QD symmetry axis, and $|\pm 2\rangle \equiv |\pm Z\rangle$. Blue upwards arrows represent pulsed excitation of the DE to its BiE where $\Omega_R(t)$ ($\Omega_L(t)$), stands for right (left) hand circularly polarized $\pi$-area pulse. Red downwards arrows denoted by $\gamma_d = 1/t_{\rm rad}$ represent spontaneous decay of the BiE to the DE associated with the emission of a photon. The coherent evolution of the DE (BiE) is represented by curved arrows denoted $\omega_2$ ($\omega_3$). The DE dephasing rate is denoted by $\gamma_2^* = 1/T_2^*$. The numerical values of the parameters used are listed in Table S1.

**Table S1.** Physical values used in our model calculations

| Quantity | Symbol | Value | Units |
| --- | --- | --- | --- |
| Biexciton lifetime[4] | $t_{\rm rad} = 1/\gamma_d$ | 0.330 | nsec |
| Biexciton precession rate[4] | $\omega_3$ | $2\pi/5.05$ | rad/nsec |
| Dark exciton precession rate[4] | $\omega_2$ | $2\pi/3.03$ | rad/nsec |
| Dark exciton dephasing time[4] | $T_2^* = 1/\gamma_2^*$ | 100 | nsec |
| DE to BiE excitation pulse width | $\hbar\Omega_W$ | 100 | μeV |
| DE to BiE excitation pulse detuning | $\Delta$ | 0 | μeV |
| DE to BiE excitation pulse area | $\Theta$ | $\pi$ | rad |
| DE pure state initialization probability | $p$ | 0.72 | |

we used the measured lower bound of Ref.[4] Since the radiative BiE to DE decay time is much shorter than the DE and BiE dephasing times, we neglected the BiE dephasing.

### 4.4 Combination of Cycles

In order to generate the multi photon entangled state, we periodically repeat the same cycle several times. In each cycle we add a new photon, entangled to its predecessor and to the DE. In modelling this procedure, the input to the $n$ cycle is the density matrix $\hat{\rho}_{n-1}$, at the end of the previous cycle. The output at the end of the cycle is $\hat{\rho}_n$, describing an $n-1$ photon and DE state. The process starts with the initial density matrix $\hat{\rho}_0 = \hat{\rho}_{\rm DE}^{\rm init}$, as described above.

We use the same physical values for each one of the cycles in our model calculation. These values, also mentioned in Fig. S5, are listed and referenced in Table S1. The model allows calculations of cases in which the polarization of the exciting pulse $\hat{\varepsilon}_{n,R/L}$, and the free evolution duration after the $n$-th pulse $\Delta t_n$ vary. These facilitate direct comparison with the experimental measurements, in which by tuning these values, we could calculate generation of a multi partite entangled state of an arbitrary length.

# 5 The Quantum Process Map

The protocol implemented in the experiment can be described by quantum circuit,[11] which is comprised of a series of cycles. The evolution of the system in each cycle is identical. Below, we first (Sec. 5.1) discuss an idealized version of the protocol, in which each cycle is comprised of unitary operations acting on the DE's spin and on the photon emitted in this cycle.

To describe the evolution of the system under the actual experimental conditions, we represent each cycle by a non-unitary quantum process map: a completely positive, trace preserving map acting on the DE's spin and on the photon emitted in this cycle. In Sec. 5.2 we define the quantum process map. In Sec. 5.3 we discuss how this process map can be determined by performing quantum process tomography.[12,13]

## 5.1 The Idealized Protocol

In an idealized version of the current experiment, each cycle of the protocol consists of a unitary transformation, as described in Fig. S6. In this figure, $|0\rangle_{\text{DE}} = |+Z\rangle$ and $|1\rangle_{\text{DE}} = |-Z\rangle$ represent the states of the dark exciton and $|0\rangle_n = |R\rangle_n$ and $|1\rangle_n = |L\rangle_n$ represent the right and left handed circular polarization states of the $n$-th photon. Note that for convenience, we use a convention in which the photon qubits are initialized in the state $|0\rangle_n$ (before they were actually emitted from the source) and the DE starts in the state $|0\rangle_{\text{DE}}$. The initialization of the DE is represented by the unitary gate $\hat{U}$ such that $|\psi_{\text{DE}}^{\text{init}}\rangle = \hat{U}|0\rangle_{\text{DE}}$. The entangling CNOT gate between the DE and the $n$-th photon is given by $\hat{C}_{\text{NOT}} = |0\rangle\langle 0|_{\text{DE}} \otimes \hat{\sigma}_{0,n} + |1\rangle\langle 1|_{\text{DE}} \otimes \hat{\sigma}_{x,n}$. Here and throughout the paper we use a notation where $\hat{\sigma}_x$, $\hat{\sigma}_y$, and $\hat{\sigma}_z$ are the Pauli matrices, while $\hat{\sigma}_0 = \mathbb{I}$ is the identity matrix. The subscript $n$, as in $\hat{\sigma}_{x,n}$, denotes that the matrices act on the $n$-th photon. The single qubit rotation $\hat{G}$, acting on the DE is a $\pi/2$ rotation around the $\hat{x}$ axis (the $\hat{x}$ direction coincides with the QD bright exciton horizontal polarization[4]), $\hat{G} = \exp\left(-i\frac{\hat{\sigma}_x}{2}\frac{\pi}{2}\right)$. A cycle in the protocol is defined as the CNOT gate applied to the DE and the emitted photon, followed by the $\hat{G}$ gate acting on the DE (marked with a dashed box in the circuit diagram Fig S6). The resulting state after $N$ cycles is identical (up to single qubit unitary transformations) to a one dimensional cluster state.[14] This cluster state is depicted on the right hand side of the circuit in Fig. S6, where open circle represents the single qubit state $\frac{1}{\sqrt{2}}(|0\rangle+|1\rangle)$, and the lines connecting two open circles represent the controlled-Z operation, $\hat{C}_{\text{Z}} = |0\rangle\langle 0|_{n-1} \otimes \hat{\sigma}_{0,n} + |1\rangle\langle 1|_{n-1} \otimes \hat{\sigma}_{z,n}$.

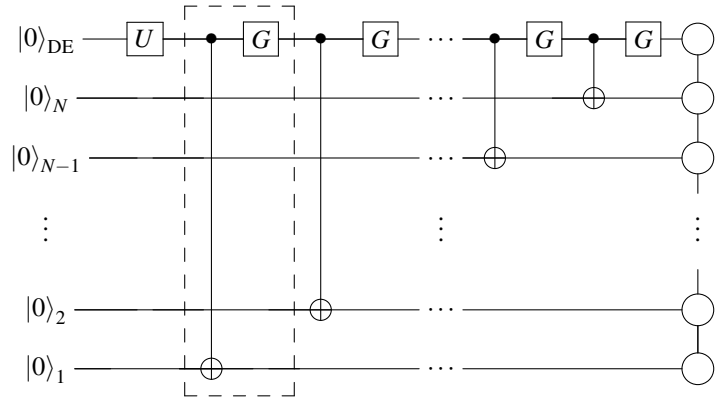

**Figure S6.** The quantum circuit representing the ideal protocol for generating a one dimensional cluster state. The unitary gates inside the dashed box represent a single cycle. The right side of the quantum circuit presents a pictorial definition of the resulting cluster state: the open circles on the right-hand side represent qubits initialized in the state $(|0\rangle+|1\rangle)/\sqrt{2}$, and the edges between them represent a controlled-Z gate.

## 5.2 Non-unitary Quantum Process Map

The physical properties of the system used in the experiment cause a deviation from the idealized unitary case discussed above. As a result, the evolution of the system in each cycle does not correspond to a unitary operation. Instead, we describe it by a completely positive trace preserving map $\Phi$. In the quantum circuit representation of the protocol, Fig. S6, this map acts non-trivially only on the degrees of freedom of the DE and the photon emitted in the cycle. Using this, and the fact that the photon qubits in the quantum circuit representation are always initialized in the fiducial state $|0\rangle_n$, allows us to consider a compact form of the process map. In this compact form, the map $\Phi$ is considered as a map from the space of $2\times 2$ density matrices describing the state of the DE spin to the space of $4\times 4$ density matrices describing the state of the DE spin and an

emitted photon. In a mathematical language, we consider a completely positive trace preserving (CPTP) map,

$$\Phi : \mathbb{C}^{2\times 2} \to \mathbb{C}^{4\times 4}. \tag{2}$$

for which $\text{Tr}[\Phi(\hat{\rho})] = 1$.

We now describe the representation we use for the map $\Phi$ throught the paper. Recall the that the DE density matrix can be represented by $\hat{\rho}_{\text{DE}} = \sum_\alpha \rho_\alpha^{(\text{DE})} \hat{\sigma}_\alpha$ with $\alpha = 0, x, y, z$ where $\hat{\sigma}_0 = \mathbb{I}_{2\times 2}$ is the identity matrix and $\hat{\sigma}_{x,y,z}$ are the $2 \times 2$, Pauli matrices. Likewise, the density matrix of the DE and an emitted photon can be decomposed as $\hat{\rho}_{\text{DE}-1} = \sum_{\alpha\beta} \rho_{\alpha\beta}^{(\text{DE}-1)} \hat{\sigma}_\alpha^{(\text{DE})} \otimes \hat{\sigma}_\beta^{(1)}$. Here, we used $\hat{\sigma}^{(\text{DE})}$ to denote Pauli matrices acting on the DE qubit, and $\hat{\sigma}^{(1)}$ to denote Pauli matrices acting on the photon qubit. The map $\Phi$ can be represented using 64 real parameters $\Phi_{\alpha\beta}^\gamma$, with $\gamma = 0, x, y, z$ representing the initial state space and $\alpha, \beta = 0, x, y, z$ representing the final state space. Its action is given by

$$\Phi(\hat{\rho}_{\text{DE}}) = \sum_{\alpha\beta} \left[\sum_\gamma \Phi_{\alpha\beta}^\gamma \rho_\gamma^{(\text{DE})}\right] \hat{\sigma}_\alpha^{(\text{DE})} \otimes \hat{\sigma}_\beta^{(1)}. \tag{3}$$

This is the representation used in Fig. 2 of the main text. Using Eq. (3) the state after $N$ application of the map can be worked out straightforwardly. For example, after two applications of the cycle, we obtain the state of the DE and two photons given by

$$\hat{\rho}^{(\text{DE}-2-1)} = \sum_{\alpha\beta\gamma} \left[\sum_{\mu\nu} \Phi_{\alpha\beta}^\nu \Phi_{\nu\gamma}^\mu \rho_\mu^{(\text{DE})}\right] \hat{\sigma}_\alpha^{(\text{DE})} \otimes \hat{\sigma}_\beta^{(2)} \otimes \hat{\sigma}_\gamma^{(1)}. \tag{4}$$

Note that in Eq. (4), $\hat{\sigma}_\alpha^{(\text{DE})}$ acts on the DE's spin, while $\hat{\sigma}_\beta^{(2)}$ and $\hat{\sigma}_\gamma^{(1)}$ act on the 2nd and 1st emitted photon, respectively.

So far, we have shown that $\Phi$ can be represented using the 64 real numbers $\Phi_{\alpha\beta}^\gamma$. However, not all possible values of $\Phi_{\alpha\beta}^\gamma$ correspond to a physical (completely positive and trace preserving) map. To find the requirements on $\Phi_{\alpha\beta}^\gamma$ which lead to a CPTP map, it is convenient to use an equivalent representation of $\Phi$ using its *Choi matrix* $C^\Phi$. The Choi matrix is defined by[15]

$$C^\Phi = \sum_{ij} e_{ij} \otimes \Phi(e_{ij}), \tag{5}$$

where $e_{ij} \in \mathbb{C}^{2\times 2}$ is 1 in the $ij$-th entry and 0 otherwise. Therefore, the Choi matrix $C^\Phi$ can be written as

$$C^\Phi = \sum_{ij} e_{ij} \otimes \left(\sum_{\alpha\beta}\sum_\gamma \Phi_{\alpha\beta}^\gamma \Lambda_{ij}^\gamma \hat{\sigma}_\alpha \otimes \hat{\sigma}_\beta\right), \tag{6}$$

where the matrices $\Lambda_{ij}^\alpha$, with $i, j \in \{1, 2\}$ and $\alpha \in \{0, x, y, z\}$ are the unitary transformations from the basis of the Pauli matrices $\hat{\sigma}_\alpha$ to the matrices $e_{ij}$, given by

$$\Lambda_{ij}^\alpha = \begin{pmatrix} 1 & 0 & 0 & 1 \\ 0 & 1 & i & 0 \\ 0 & 1 & -i & 0 \\ 1 & 0 & 0 & -1 \end{pmatrix}. \tag{7}$$

Choi's theorem[15] states that $\Phi$ is a completely positive map, if and only if its Choi matrix $C^\Phi$ is a positive matrix. Therefore, Choi's theorem and Eq. (6) give (equivalent) conditions on the values of $\Phi_{\alpha\beta}^\nu$ such that they correspond to a completely positive map $\Phi$. The requirement that $\Phi$ is trace preserving, $\text{Tr}[\Phi(\hat{\rho})] = 1$, gives, using Eq. (3), the following conditions

$$\Phi_{00}^0 = 1/2; \qquad \Phi_{00}^\gamma = 0, \gamma \neq 0. \tag{8}$$

The conditions in Eq. (8) lead to the following conditions on $C^\Phi$,

$$\text{Tr}\left[C^\Phi \hat{\sigma}_0 \otimes \hat{\sigma}_0 \otimes \hat{\sigma}_0\right] = 2; \qquad \text{Tr}\left[C^\Phi \hat{\sigma}_\alpha \otimes \hat{\sigma}_0 \otimes \hat{\sigma}_0\right] = 0, \alpha \neq 0. \tag{9}$$

### 5.3 Quantum Process Tomography of the Process Map

The process map can be obtained experimentally by performing quantum process tomography. This is achieved using the following three steps:

### 5.3.1 Step 1
The DE is prepared in four linearly independent initial states $\hat{\rho}_{DE}(j)$, $j=1,....,4$. Full quantum state tomography on the final density matrix of the DE and the emitted photon leads to the following set of equations

$$\rho^{(DE-1)}_{\alpha\beta}(j) = \sum_{\gamma} \Phi^{\gamma}_{\alpha\beta} \rho^{(DE)}_{\gamma}(j). \tag{10}$$

Due to the selection rules of the DE-BiE transitions, we cannot directly measure the DE in the $|+X\rangle$ and $|-X\rangle$ states, which means that we cannot directly measure the four elements $\rho^{(DE-1)}_{x\beta}$. Therefore, using Eq. (10) we can only obtain the 48 elements $\Phi^{\gamma}_{\alpha\beta}$ for $\alpha \neq x$.

### 5.3.2 Step 2
The elements $\Phi^{\gamma}_{x\beta}$, which were not determined by Eq. (10), can be obtained experimentally by applying two cycles of the protocol, and measuring correlations between the state of the DE and the state of the two emitted photons. We consider the state after two cycles, given four linearly independent initial states $\hat{\rho}^{(DE)}(j)$, $j=1,....,4$ of the DE. As before, we expand the state of the DE and the two emitted photons as $\hat{\rho}^{(DE-2-1)} = \sum_{\alpha\beta\gamma} \rho^{(DE-2-1)}_{\alpha\beta\gamma} \hat{\sigma}^{(DE)}_{\alpha} \otimes \hat{\sigma}^{(2)}_{\beta} \otimes \hat{\sigma}^{(1)}_{\gamma}$. Using Eq. (4) we get the set of equations:

$$\rho^{(DE-2-1)}_{\alpha\beta\gamma}(j) = \sum_{\mu\nu} \Phi^{\nu}_{\alpha\beta} \Phi^{\mu}_{\nu\gamma} \rho^{(DE)}_{\mu}(j). \tag{11}$$

This equation can be rewritten as

$$\rho^{(DE-2-1)}_{\alpha\beta\gamma}(j) - \sum_{\mu}\sum_{\nu \neq x} \Phi^{\nu}_{\alpha\beta} \Phi^{\mu}_{\nu\gamma} \rho^{(DE)}_{\mu}(j) = \sum_{\mu} \Phi^{x}_{\alpha\beta} \Phi^{\mu}_{x\gamma} \rho^{(DE)}_{\mu}(j). \tag{12}$$

Let us consider a subset of the equations in (12) for specific values of $\alpha, \beta$ and $\gamma$, such that $\alpha \neq x$. Importantly, for this case all of the elements in the first term on the left hand side of Eq. (12) can be obtained experimentally, by performing measurements after application of *two* cycles of the protocol. All the elements of the second term were obtained in in step 1 [i.e., by performing measurements after application of a *single* cycle and by using Eq. (10)]. On the right hand side, the elements $\Phi^{\mu}_{x\gamma}$ are the unknowns, but the elements $\Phi^{x}_{\alpha\beta}$ were obtained in step 1 (recall that $\alpha \neq x$). Therefore, if $\Phi^{x}_{\alpha\beta} \neq 0$ we obtain a set of 16 linear equations for the 16 missing elements $\Phi^{\mu}_{x\gamma}$, which completes the quantum process tomography.

The total set of 64 equations consisting of 48 equations from step 1 and the 16 equations from step 2 corresponds to a complete set of measurements, which uniquely determines $\Phi^{\gamma}_{\alpha\beta}$.

### 5.3.3 Step 3
In the actual experiment, the tomography is obtained from a finite number of 2 or 3 correlated events, in which single photons are timely recorded by 2 or 3 detectors, respectively. This invariably leads to statistical errors in the measured expectation values. Therefore, the process map obtained by solving Eqs. (10) and (12) using the measured values directly is not necessarily a physical process map, i.e., the corresponding Choi matrix $C^{\Phi}$ can have negative eigenvalues, or Eq. (9) may not be obeyed. Therefore, in the final step of the quantum process tomography, we look for a physical process map which best fits our experimental measurements. To achieve this, we use an algorithm which randomly samples the space of physical process maps, and searches for the physical process map which minimizes the distance to the measured process map. Using this sampling procedure, we achieve simultaneously two goals. The first goal is to find the physical process map which best fits the measured map within the experimental uncertainties. The second goal is to verify that this physical process map is uniquely determined by our measurements.

The sampling procedure employed by the algorithm proceeds as follows. Starting from an initial process map $\Phi_0$, which is represented by an $8 \times 8$ Choi matrix $C^{\Phi_0}$, the algorithm chooses a random Hermitian $8 \times 8$ matrix $H^{\Phi}$, by expanding $\log(H^{\Phi})_{\alpha\beta\gamma} = \sum_{\alpha\beta\gamma} L^{\Phi}_{\alpha\beta\gamma} \hat{\sigma}_{\alpha} \otimes \hat{\sigma}_{\beta} \otimes \hat{\sigma}_{\gamma}$ and drawing the coefficients $L^{\Phi}_{\alpha\beta\gamma}$ according to the distribution

$$\mu_{\Phi_0,\Sigma}\left(L^{\Phi}_{\alpha\beta\gamma}\right) = \frac{1}{\sqrt{2\pi\Sigma^2}} \exp\left[-\frac{(L^{\Phi}_{\alpha\beta\gamma} - L^{\Phi_0}_{\alpha\beta\gamma})^2}{2\Sigma^2}\right]. \tag{13}$$

where $L^{\Phi_0}_{\alpha\beta\gamma}$ are the expansion coefficients of $\log(C^{\Phi_0})$ in the Pauli basis. Here, the subscripts $\alpha\beta\gamma$ imply that the drawing is done in the Pauli basis, in which all the coefficients are real. The width of the distribution $\Sigma$ is randomly chosen between 0.03

and 300 using a uniform distribution on a $\log_{10}$ scale. Small values of $\Sigma$ imply initial conditions which are close to $C^{\Phi_0}$, while large values imply distant conditions. The matrix $H = \exp(L)$ is Hermitian and positive. In order to yield a physical Choi matrix $C^\Phi$, the matrix $H^\Phi$ is then normalized according to

$$C^\Phi = \left(S_1^\Phi \otimes \mathbb{I}_{4\times 4}\right) \bar{H}^\Phi \left(S_1^\Phi \otimes \mathbb{I}_{4\times 4}\right), \tag{14}$$

where $\bar{H}^\Phi = 2H^\Phi/\text{Tr}[H^\Phi]$ is the normalized Hermitian matrix, and the multiplication by $S_1^\Phi = \left(\text{Tr}_{2,3}[\bar{H}^\Phi]\right)^{-1/2}$ ensures that Eq. (9) is obeyed (here $\text{Tr}_{2,3}$ refers to the partial trace of the 2nd and 3rd "qubits" of the Choi matrix, as presented in Sec. 5.2).

To quantify how well the process map $\Phi$ reconstructs the complete set of measurements (see step 2 above), we compute a residue function $R(\Phi)$, which measures the weighted distance from the measured values. This function will be defined in Sec. 5.3.4 below. We then numerically minimize $R(\Phi)$: starting from $C^{\Phi_0}$, we use a Nelder-Mead simplex search algorithm to find a map $\Phi_{\min}$, which gives a local minimum of the function $R(\Phi)$. The converged value $\Phi_{\min}$ depends mainly on the reference point $\Phi_0$, and the distribution width $\Sigma$.

We now iterate the procedure by the following Monte Carlo approach. In the beginning of each iteration, we first consider the starting point $\Phi_0$ and end point $\Phi_{\min}$ of the previous iteration. We then set a new value for $\Phi_0$ as $\Phi_0 = \Phi_{\min}$ if $R(\Phi_{\min}) < R(\Phi_0)$; otherwise, we keep the same $\Phi_0$. We then choose a new value for $\Sigma$ and draw a Hermitian matrix $H^\Phi$ using the distribution $\mu_{\Phi_0,\Sigma}$ of Eq. (13). Next, we compute the corresponding $C^\Phi$ using Eq. (14). We conclude each iteration by using $C^\Phi$ as the starting point for the Nelder-Mead algorithm for finding $\Phi_{\min}$ which gives a local minimum of $R(\Phi)$.

After $10^4$ iterations, we obtain a large set of process maps $\{\Phi_{\min}\}$. The results are visualized in Fig. S7 where for each $\Phi_{\min}$ we plot $R(\Phi_{\min})$ versus the distance between $\Phi_{\min}$ and the process map obtained from the unitary process map of the ideal protocol. The latter distance is quantified by the fidelity between the respective Choi matrices $\mathscr{F}\left(C_{\min}^\Phi, C_{\text{ideal}}^\Phi\right)$ (see Sec. 9 below for the definition of $\mathscr{F}$).

In Fig. S7, we show that the two goals we set were obtained. First, we obtain the physical process map which best fits our complete set of measurements. The best fit is quantified by finding a global minimum for the residue $R(\Phi)$, which is well within the experimental uncertainties (horizontal dashed line in Fig. S7). Second, Fig. S7 shows that this global minimum is unique. The process map which gives this global minimum, and therefore gives the best fit to the measurements, is marked by a black triangle and denoted by $\Phi_{\text{measured}}$. This process map is shown in Fig. 2(a) of the main text. It has excellent similarity to the ideal process map, as defined by the fidelity between their Choi matrices $\mathscr{F}\left(C_{\text{measured}}^\Phi, C_{\text{ideal}}^\Phi\right) = 0.811 \pm 0.020$ (see Sec. 9 below for the definition of $\mathscr{F}$).

### 5.3.4 The Residue Function
The residue function, which estimates the deviation between a drawn physical process map and the measured map, is given by

$$R(\Phi) = \sqrt{\frac{\sum_{k=1}^{64} n_k (P_k^\Phi - P_k^{\text{meas}})^2}{\sum_{k=1}^{64} n_k}}, \tag{15}$$

where the sum runs over all 64 independent correlation measurements described in step 1 and step 2 above. Each term in the sum corresponds to a specific experimental configuration $k$, where each configuration consists of one out of four different polarizations of the initial DE state, one out of four different polarizations for each of the emitted photons and one out of four different final polarizations of the DE. We denote by $n_k$ the number of measured correlation events in the configuration $k$, and by $P_k^{\text{meas}}$ and $P_k^\Phi$ the measured and calculated probability of detecting this particular configuration, respectively. The calculated probability is computed from the process map $\Phi$, as described in sections step 1 and step 2 above. For example, suppose that the $k$ configuration corresponds to preparation of the DE in the $j_0$ state, and after one cycle the DE is measured in spin state $|+Y\rangle$ and the emitted photon in polarization state $|H\rangle$. Then $P_{+Y,H}^\Phi = \langle +Y, H | \hat{\rho}^{(\text{DE}-1)}(j_0) | +Y, H \rangle$, where $\hat{\rho}^{(\text{DE}-1)}(j_0)$ is computed from $\Phi$ according to Eq. (10). The measured probability $P_k^{\text{meas}}$ is obtained by dividing $n_k$ by a normalization constant, which is calculated by summing over all correlated events using two orthogonal polarizations for each detected photon or spin state of the DE. For example, for the configuration used in the example above, we get $P_{+Y,H}^{\text{meas}} = n_{+Y,H} / (n_{+Y,H} + n_{+Y,V} + n_{-Y,H} + n_{-Y,V})$ The function $R(\Phi)$ therefore quantifies how much the process map $\Phi$ deviates from the measured data, where the measured 64 different experimental configurations are weighted by their statistical significance.

## 6 Localizable Entanglement

The quantification of entanglement in multiparticle systems is important, as it evaluates the amount of quantum information processing (QIP) tasks that can be performed on the system. Localizable entanglement (LE)[16] is a measure of this multiparticle entanglement.

In Sec. 6.1 we define the LE. In Sec. 6.2, we apply this definition to our experimental results.

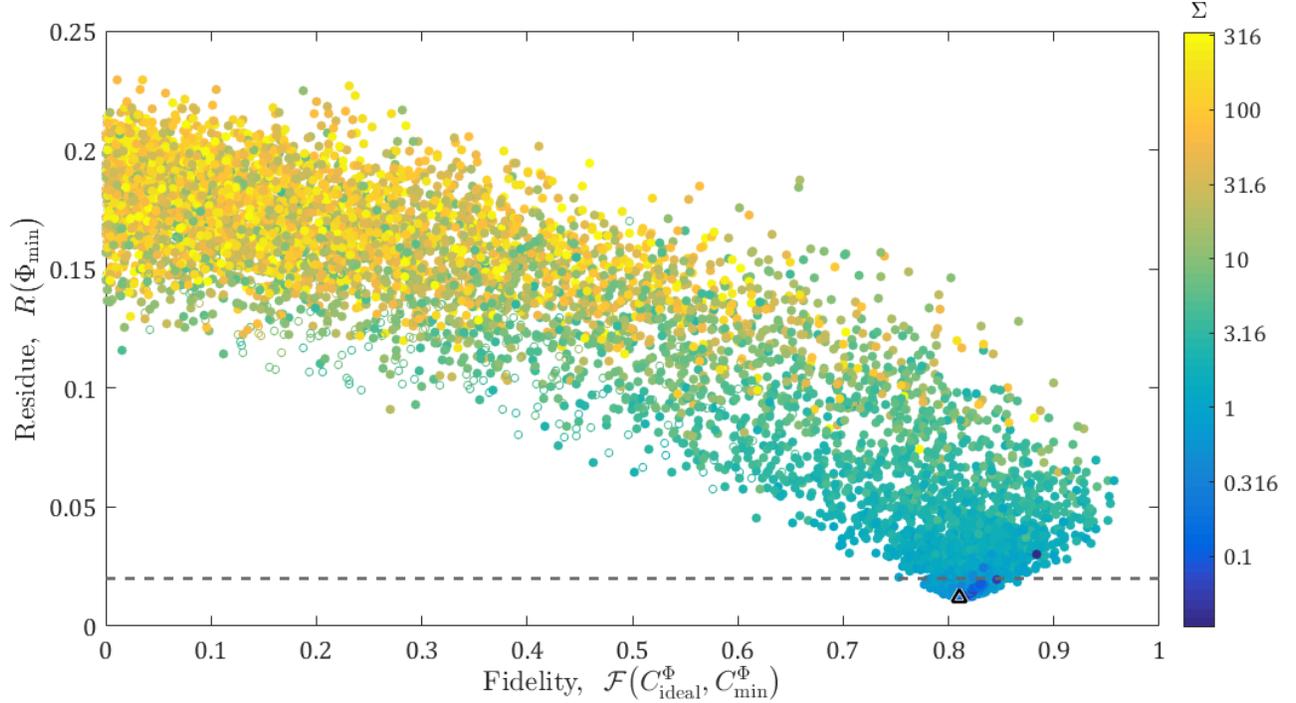

**Figure S7.** Monte Carlo results. Each point corresponds to a process map $\Phi_{\min}$ obtained at the end of one iteration. The points indicate the residue $R(\Phi_{\min})$ versus the fidelity between their corresponding Choi matrix and the Choi matrix of the ideal protocol. The color represents the width $\Sigma$ of the distribution in Eq. (13), used in each iteration. A filled point depicts a positive localizable entanglement (see Sec. 6) between the two photons in the 3 qubit density matrix when the DE spin is projected on $|+Z\rangle$ state. An empty point means no entanglement. The black triangle depicts the global minimum, denoted by $\Phi_{\text{measured}}$, which represents the process map that fits best to the experimental measurements. The dashed gray line represents the experimental uncertainties at $R(\Phi_{\min}) = 0.02$.

### 6.1 Definition

Given a state $\hat{\rho}_N$ of $N$ qubits, we define the LE[17] of two qubits $n$ and $m$ by the maximal negativity $\mathcal{N}$ of their reduced density matrix obtained after performing projective single qubit measurements on all the other $N-2$ qubits. We denote by $\mathcal{M}$ a choice for the basis for the projective measurements, and by $s$ the outcomes of the measurements. For a specific $\mathcal{M}$ and $s$, the reduced density matrix between qubits $n$ and $m$ is given by

$$\hat{\rho}_{n,m}^{(\mathcal{M},s)} = \frac{1}{p_s} \text{Tr}_{j\neq m,n}\left[\left(\prod_{j\neq m,n} \hat{P}_j\right) \hat{\rho}_N\right]. \tag{16}$$

In the above, the basis $\mathcal{M}$ and the outcomes $s$ correspond to the sequence of projections $\prod_{j\neq m,n}\hat{P}_j$, where $\hat{P}_j = |s_j\rangle\langle s_j|$ is a projector acting on qubit $j$. Note that all qubits are projected except for qubits $n$ and $m$. Likewise, the partial trace is taken over all qubits except qubits $m$ and $n$. Finally, $p_s$ is the probability to obtain the measurement outcome $s$, given by $p_s = \text{Tr}\left[\prod_{j\neq m,n}\hat{P}_j\hat{\rho}_N\right]$. The average localizable entanglement between the qubits $n$ and $m$, maximized over all possible measurement basis is then given by[17]

$$\mathcal{N}_{n,m}^{\text{LE}}(\hat{\rho}_N) = \max_{\mathcal{M}} \sum_s p_s \mathcal{N}\left(\hat{\rho}_{n,m}^{(\mathcal{M},s)}\right). \tag{17}$$

where $\mathcal{N}$ is the negativity of the density matrix $\hat{\rho}_{n,m}$. The negativity is defined as follows. Consider a density matrix describing the state of two qubits $\hat{\rho} = \sum \rho_{m\mu;n\nu}|m\rangle\langle n|\otimes|\mu\rangle\langle\nu|$, where the labels $m,n$ correspond to qubit $A$, and $\mu,\nu$ belong to qubit $B$, and the sum is over the labels $m,n,\mu,\nu = 0,1$. The partial transpose of $\hat{\rho}$ is defined by $\hat{\rho}^{T_P} = \sum \rho_{m\mu;n\nu}|m\rangle\langle n|\otimes(|\mu\rangle\langle\nu|)^T = \sum \rho_{m\nu;n\mu}|m\rangle\langle n|\otimes|\nu\rangle\langle\mu|$. The state $\hat{\rho}$ is entangled if and only if any of the eigenvalues of $\hat{\rho}^{T_P}$ is negative.[18,19] Denoting by $\lambda_j$ the 4 eigenvalues of $\hat{\rho}^{T_P}$, the negativity[20] $\mathcal{N}$ is defined as $\mathcal{N}(\hat{\rho}) = \frac{1}{2}\left(\sum_j|\lambda_j| - \lambda_j\right)$, i.e., $\mathcal{N}(\hat{\rho})$ is the sum of the absolute

value of the negative eigenvalues of $\hat{\rho}^{T_P}$. The negativity measures the degree of entanglement in $\hat{\rho}$, where $\mathcal{N} = 0$ corresponds to an unentangled (separable) state, and $\mathcal{N} = \frac{1}{2}$ to a maximally entangled state.

## 6.2 Localizable Entanglement in the Quantum State Generated in the Experiment

In an ideal cluster state, the LE between any two qubits is maximal, $\mathcal{N}_{n,m}^{LE} = 1/2$. It is convenient to give a position label to the qubits in the quantum state $\hat{\rho}_{N+1}$ generated after applications of $N$ cycles of the protocol, such that the qubit at position $N+1$ is the DE, the first emitted photon is at position 1, the second emitted photon is at position 2, and the $n$-th emitted photon is at position $n$ (see Fig. 1(d) of the main text). For the states $\hat{\rho}_{N+1}$ generated in the experiment, the LE will depend on the distance $d$ between two qubits in the chain. If the number of qubits $N+1$ is sufficiently larger then $d$, the LE *will not* depend on $N$ (for $N \approx d$ the LE can show a weak dependence on $N$). We define the *entanglement length*[16] $\xi_{LE}$, which measures the decay length scale of the LE.

To obtain the LE in the state generated in the experiment, we compute the state $\hat{\rho}_{N+1}$ by applying the process map $\Phi_{\text{measured}}$ to the initial partially mixed DE state $\hat{\rho}_{DE}^{\text{init}} = 0.72 |-X\rangle\langle -X| + 0.28 \mathbb{I}$. Using Eqs. (16) and (17), we then compute the expected localizable entanglement between the DE's spin and the first emitted photon. These two qubits are at a distance $d = N$ in terms of the position labels discussed above. The resulting expected localizable entanglement, $\mathcal{N}_{1,N+1}^{LE}$ is shown as the blue points in Fig. 3 in the main paper. To obtain an error estimate on the LE, we consider all process maps $\Phi_{\text{min}}$ obtained in Sec. 5.3.3 with a residue of $R(\Phi_{\text{min}}) < 0.02$, which is compatible with our experimental uncertainties.

The localizable entanglement persists across 5 qubits, and decays as $\mathcal{N}_{1,N+1}^{LE} = \mathcal{N}_0 \exp(-N/\xi_{LE})$ with $\xi_{LE} = 3.2$. The localizable entanglement between two qubits at a distance $d$ in the *interior* of long chains with $N \gg d$ is in fact independent of the purity of the initial state of the DE.[11]

# 7 Direct Measurement of Two Qubit Entanglement

In this section we describe direct verification of entanglement between each pair of qubits in our three qubit chain. In Sec. 7.1 we discuss entanglement between the first two photons, conditioned on measuring the DE. In Sec. 7.2 we discuss entanglement between the DE and the nearest-neighboring photon, and in Sec 7.3 we discuss the localizable entanglement between the DE and the next-nearest neighboring photon conditioned on measuring the polarization of the photon in-between.

## 7.1 Photon–Photon Localizable Entanglement

We conducted four sets of 16 polarization-projective measurements. In each set we initialized the DE in its $|\Psi_{DE}^{\text{init}}\rangle = |-X\rangle$ eigenstate, and then used two timed $|H\rangle$ polarized pulses, corresponding to application of two cycles of the protocol, and a third readout pulse in four different configurations, as described in the method section of the main paper. As a result of the pulse sequence, three photons are emitted, their correlated detection comprised a measured event. The last pulse was polarized in either $|R\rangle$ or $|L\rangle$, with appropriate timing as to project the DE spin on the $|\pm Z\rangle$ or $|\pm Y\rangle$ states. In each case, the first two photons are measured using 16 different mutual polarization projections, from which we reconstruct the density matrix of the two photons. Specifically, each photon was projected on the $|H\rangle$, $|V\rangle$, $|L\rangle$, and $|D\rangle$ polarizations. For example, the density matrix element $\rho_{LL,LL}$ in the circular polarization basis was obtained by $\rho_{LL,LL} = n_{LL}/(n_{HH} + n_{HV} + n_{VH} + n_{VV})$ where $n_{P_1P_2}$ is the measured number of correlated photons in the two photon $|P_1P_2\rangle$ polarization configuration.

The density matrices we obtained from these polarization tomographies[21] are presented in Fig. S8. For reasons of simplicity in the presentation of these matrices the first photon is always displayed in the circular polarization basis while the second photon is displayed in a linear polarization basis. In the figure, we denote the density matrices by $\hat{\rho}_{P_{DE}}^{\text{meas}}$, where $P_{DE}$ corresponds to the state onto which the DE was projected. Only for the first two cases Fig. S8(a) and (b), corresponding to projections of the DE onto the states $|R\rangle$ and $|L\rangle$ the density matrices are expected to correspond to an entangled state. Indeed, their negativities $\mathcal{N}$ are $0.242 \pm 0.052$ and $0.248 \pm 0.069$, and their fidelities $\mathcal{F}$ to the appropriate Bell state (see Fig. S8) are $0.730 \pm 0.060$ and $0.722 \pm 0.054$, respectively. These values show entanglement between the two photons and serve as a measure of the localizable entanglement between adjacent qubits, as discussed in Sec. 6 above and presented in Fig. 3 of the main paper. The fidelities of the density matrices in Fig. S8(c) and (d) to the appropriate pure state, as expected from the ideal protocol, are $0.633 \pm 0.037$ and $0.681 \pm 0.048$, respectively. The "expected localizable entanglement" between the two photons (after optimizing the measurement basis and averaging over the measurement outcomes) gives

$$\mathcal{N}_{2-1}^{LE} = 0.168. \tag{18}$$

## 7.2 DE–Photon Entanglement

A direct measurement of the density matrix of the DE and the emitted photon (after application of a single cycle) requires full state tomography of both qubits. However, projections of the DE on the states $|\pm X\rangle$, which are required for full tomography,

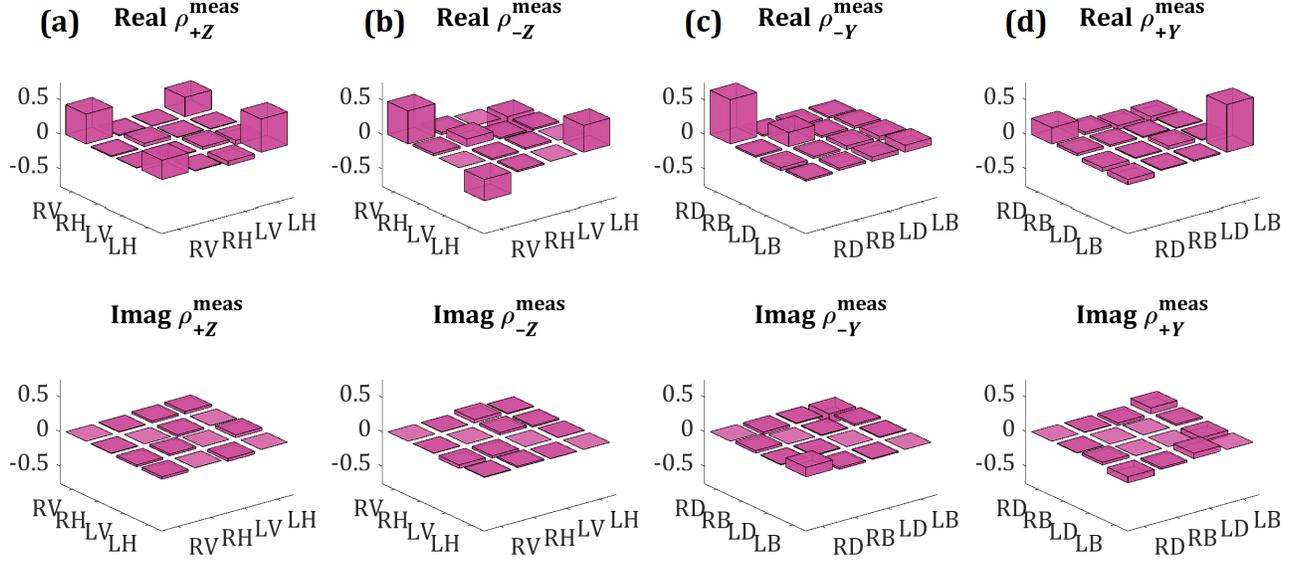

**Figure S8.** a) [b)] The measured two-photon density matrix obtained by polarization tomography of the two emitted photons conditioned on detecting the third photon, following a right [left] hand circularly polarized DE projection pulse, 3/4 of a precession period after the second converting pulse. This detection, therefore projects the DE on the $|+Z\rangle$ $[|-Z\rangle]$ state. c) [d)] The measured two-photon density matrix as in a) [b)], but this time, the circularly polarized DE projection pulse is timed 1/2 a precession period after the second conversion pulse. This way the DE is projected on the $|-Y\rangle$ $[|+Y\rangle]$ state. In all cases the DE is initialized in the $|-X\rangle$ state, and the first photon is displayed in the circular polarization basis. The second photon in a) and b) is displayed in the rectilinear polarization basis and in c) and d) in the linear diagonal polarization basis. The fidelities to Bell-states $(|RV\rangle + |LH\rangle)/\sqrt{2}$ in a) and to $(|RV\rangle - |LH\rangle)/\sqrt{2}$ in b) are $0.730 \pm 0.060$ and $0.722 \pm 0.054$, respectively. The fidelities to the pure states $|RD\rangle$ in c) and $|LB\rangle$ in d) are $0.633 \pm 0.037$ and $0.681 \pm 0.048$, respectively.

are not straightforward and require additional resources. Fortunately, projective measurements on the $|\pm Z\rangle$ and $|\pm Y\rangle$ bases are sufficient to provide a lower bound[22] on the fidelity between the DE-photon density matrix and the maximally entangled state of two qubits,

$$|\psi\rangle_{DE-1} = \left[(|+Z\rangle + i|-Z\rangle)|R_1\rangle - (|-Z\rangle + i|+Z\rangle)|L_1\rangle\right]/2. \tag{19}$$

The fidelity between the DE-photon density matrix, $\hat{\rho}^{(DE-1)}$ and the state $|\psi\rangle_{DE-1}$ is defined as

$$\mathscr{F}\left(\hat{\rho}^{(DE-1)}, |\psi\rangle\langle\psi|_{DE-1}\right) = {}_{DE-1}\langle\psi|\hat{\rho}^{(DE-1)}|\psi\rangle_{DE-1}. \tag{20}$$

The lower bound on $\mathscr{F}$ can be written as[22]

$$\mathscr{F}\left(\hat{\rho}^{(DE-1)}, |\psi\rangle\langle\psi|_{DE-1}\right) \geq (\mathscr{F}_1 + \mathscr{F}_2)/2. \tag{21}$$

where $\mathscr{F}_1$ and $\mathscr{F}_2$ are defined as

$$\mathscr{F}_1 = \rho^{(DE-1)}_{ZR,ZR} + \rho^{(DE-1)}_{-ZL,-ZL} - \sqrt{\rho^{(DE-1)}_{ZL,ZL} \cdot \rho^{(DE-1)}_{-ZR,-ZR}}, \tag{22}$$

and

$$\mathscr{F}_2 = \rho^{(DE-1)}_{YB,YB} + \rho^{(DE-1)}_{-YD,-YD} - \left(\rho^{(DE-1)}_{YD,YD} + \rho^{(DE-1)}_{-YB,-YB}\right). \tag{23}$$

We experimentally measured both $\mathscr{F}_1$ and $\mathscr{F}_2$. To obtain $\mathscr{F}_1$, we measured the DE in the $|\pm Z\rangle$ basis and the photon in the $|R\rangle$ and $|L\rangle$ basis. For $\mathscr{F}_2$, we measured the DE in the $|\pm Y\rangle$ basis and the photon in the $|D\rangle$ and $|B\rangle$ bases. We got

$$\mathscr{F}\left(\hat{\rho}^{(DE-1)}, |\psi\rangle\langle\psi|_{DE-1}\right) \geq 0.705 \pm 0.041. \tag{24}$$

The lower bound on the entanglement fidelity that we obtained this way is similar to that obtained in Refs.[22–26].

Importantly, the state $\hat{\rho}^{(\text{DE}-1)}$ is entangled if $\mathscr{F}(\hat{\rho}^{(\text{DE}-1)}, |\psi\rangle\langle\psi|_{\text{DE}-1}) > 0.5$. The negativity of $\hat{\rho}^{(\text{DE}-1)}$ for 2 qubits is directly related to $\mathscr{F}$ using

$$\mathscr{N}(\hat{\rho}^{(\text{DE}-1)}) \geq \mathscr{F}(\hat{\rho}^{(\text{DE}-1)}, |\psi\rangle\langle\psi|_{\text{DE}-1}) - 0.5. \tag{25}$$

Therefore, our measurements also yield a lower bound on the negativity $\mathscr{N}(\hat{\rho}^{(\text{DE}-1)}) \geq 0.205 \pm 0.041$.

### 7.3 DE–Next-nearest-photon Localizable Entanglement

We now discuss the localizable entanglement between the DE and the first emitted photon in the three qubit state obtained after applying two cycles of the protocol. The localizable entanglement measures the degree of entanglement in the density matrix between of the DE and the first emitted photon,

$$\hat{\rho}_{P_2}^{(\text{DE}-1)} = \text{Tr}_2\left[\hat{\rho}^{(\text{DE}-2-1)}(\mathbb{I} \otimes |P_2\rangle\langle P_2| \otimes \mathbb{I})\right], \tag{26}$$

obtained after a projective measurement of the second emitted photon on the state $|P_2\rangle$. Since we cannot measure the DE in the basis $|\pm X\rangle$, we used the same approach as discussed in Sec. 7.2. Choosing $|P_2\rangle = |V\rangle$ we obtained a lower bound on the fidelity of $\hat{\rho}_V^{(\text{DE}-1)}$ with the state

$$|\psi\rangle_{\text{DE}-1} = (|+ZR\rangle + |-ZL\rangle)/\sqrt{2}, \tag{27}$$

which is equivalent to a bound on the negativity of $\hat{\rho}_V^{(\text{DE}-1)}$. Note that while in Sec. 7.2 two photon correlations were sufficient, here three photon correlations were required.

The data presented in Fig. S8 contain the required information needed to obtain the lower bound. As discussed above, in Fig. S8 four density matrices of two photons are presented. Each matrix presents two photon density matrix while the DE is projected on one the states $|+Z\rangle, |-Z\rangle, |+Y\rangle$ and $|-Y\rangle$ (see Fig. S8 (a), (b), (c) and (d), respectively). Using this data, we construct the elements of the density matrix $\hat{\rho}_{P_2}^{(\text{DE}-1)}$ which are necessary for obtaining the quantities $\mathscr{F}_1$ and $\mathscr{F}_2$ defined in Eq. (22) and Eq. (23). Next, we optimize the projective measurement polarization $P_2$ in order to maximize $(\mathscr{F}_1 + \mathscr{F}_2)/2$. We find that the optimal polarization is $|\tilde{V}\rangle = \exp(i\theta\hat{\sigma}_z/2)|V\rangle$ with $\theta = -11°$, for which we obtain for the 2 remaining qubits

$$\mathscr{F}(\hat{\rho}_{\tilde{V}}^{(\text{DE}-1)}, |\psi\rangle\langle\psi|_{\text{DE},1}) \geq 0.590 \pm 0.062 \quad \Rightarrow \quad \mathscr{N}(\hat{\rho}_{\tilde{V}}^{(\text{DE}-1)}) \geq 0.090 \pm 0.062. \tag{28}$$

## 8 Direct Measurement of Genuine Three Qubit Entanglement

In this section we show that the 3 qubit state that was generated in the experiment, consisting of the DE and the two emitted photons, exhibits genuine three qubit entanglement. A three qubit state $\hat{\rho}_3$ with genuine three qubit entangled is defined as a state of three qubits, $A, B$ and $C$ which is not bi-separable, i.e., it *cannot* be written as $\hat{\rho}_3 = \sum_k p_k^{(1)} \hat{\rho}_k^{(A)} \otimes \hat{\rho}_k^{(BC)} + \sum_l p_l^{(2)} \hat{\rho}_l^{(AB)} \otimes \hat{\rho}_l^{(C)} + \sum_m p_m^{(3)} \hat{\rho}_m^{(AC)} \otimes \hat{\rho}_m^{(B)}$, where $\hat{\rho}_l^{(A)}, \hat{\rho}_l^{(B)}$ and $\hat{\rho}_l^{(C)}$ are single qubit states of the qubits $A, B$ and $C$, respectively, $\hat{\rho}_m^{(AB)}, \hat{\rho}_m^{(AC)}$ and $\hat{\rho}_m^{(BC)}$ are two qubits states of the qubits $AB, AC$ and $BC$, respectively and $p_k^{(j)} \geq 0$ with $\sum_{j,k} p_k^{(j)} = 1$.

Since we do not perform measurements of the DE in the basis $|\pm X\rangle$, we do not perform full tomography on the three qubit (DE-photon-photon) state. Instead, as in Secs. 7.3 and 7.2, we use our measurements to demonstrate that the three qubit state is genuinely three qubit entangled in two complementary ways. First, in Sec. 8.1 we obtain a lower bound on the fidelity of the density matrix of the 3 qubits with that of the three qubit cluster state. Second, in Sec. 8.2 we construct an entanglement witness and measure its expectation value. Each approach separately demonstrates that the DE-photon-photon state is genuinely three qubit entangled.

### 8.1 Fidelity to a Pure Cluster State

The fidelity between the DE-photon-photon state and the three qubit cluster state is given by

$$\mathscr{F}(\hat{\rho}^{(\text{DE}-2-1)}, |\psi\rangle\langle\psi|_{\text{C}}) = {}_{\text{C}}\langle\psi|\hat{\rho}^{(\text{DE}-2-1)}|\psi\rangle_{\text{C}}, \tag{29}$$

where $\hat{\rho}^{(\text{DE}-2-1)}$ is the state of the DE and the two emitted photons, and

$$|\psi\rangle_{\text{C}} = \Big(|+Z\rangle\big[(|R_2\rangle - |L_2\rangle)|R_1\rangle - i(|R_2\rangle + |L_2\rangle)|L_1\rangle\big] \\ + |-Z\rangle\big[i(|R_2\rangle + |L_2\rangle)|R_1\rangle + (|R_2\rangle - |L_2\rangle)|L_1\rangle\big]\Big)/2\sqrt{2}. \tag{30}$$

**Table S2.** Results of the stabilizer measurements. For each $\hat{S}_i$, we give $\hat{S}_i\hat{\rho}$ in terms of the measured density matrices, which are described in Fig. S8; $\text{Tr}_{\text{DE}}$ is a trace over the DE spin; $s_i^{\text{meas}} = \text{Tr}[\hat{S}_i\hat{\rho}]$, the measured stabilizer observable value; and $\Delta s_i$, its measurement error.

| Operator $\hat{S}_i$ | Density Matrix $\text{Tr}_{\text{DE}}[\hat{S}_i\hat{\rho}]$ | Measured Value $s_i^{\text{meas}}$ | Error Value $\Delta s_i$ |
|---|---|---|---|
| $\hat{S}_1$ | $+(\hat{\sigma}_z \otimes \hat{\sigma}_y)(\hat{\rho}_{+Z}^{\text{meas}} + \hat{\rho}_{-Z}^{\text{meas}})$ | 0.780 | 0.058 |
| $\hat{S}_2$ | $-(\hat{\sigma}_x \otimes \hat{\sigma}_z)(\hat{\rho}_{+Z}^{\text{meas}} - \hat{\rho}_{-Z}^{\text{meas}})$ | 0.619 | 0.061 |
| $\hat{S}_3$ | $-(\hat{\sigma}_y \otimes \hat{\sigma}_x)(\hat{\rho}_{+Z}^{\text{meas}} - \hat{\rho}_{-Z}^{\text{meas}})$ | 0.555 | 0.107 |
| $\hat{S}_4$ | $-(\hat{\sigma}_0 \otimes \hat{\sigma}_y)(\hat{\rho}_{+Y}^{\text{meas}} - \hat{\rho}_{-Y}^{\text{meas}})$ | 0.409 | 0.056 |
| $\hat{S}_5$ | $-(\hat{\sigma}_z \otimes \hat{\sigma}_0)(\hat{\rho}_{+Y}^{\text{meas}} - \hat{\rho}_{-Y}^{\text{meas}})$ | 0.586 | 0.032 |

Importantly, if $\mathscr{F}(\hat{\rho}^{\text{DE}-2-1}, |\psi\rangle\langle\psi|_{\text{C}}) > 0.5$ then the three qubit state is genuinely three qubit entangled.

Consider the following eight mutually commuting observables (the first one being the identity) $\hat{S}_i$, $i = 0, 1, \ldots, 7$

$$\begin{aligned}
\hat{S}_0 &= +\hat{\sigma}_0 \otimes \hat{\sigma}_0 \otimes \hat{\sigma}_0, & \hat{S}_1 &= +\hat{\sigma}_0 \otimes \hat{\sigma}_z \otimes \hat{\sigma}_y, & \hat{S}_2 &= -\hat{\sigma}_z \otimes \hat{\sigma}_x \otimes \hat{\sigma}_z, & \hat{S}_3 &= -\hat{\sigma}_z \otimes \hat{\sigma}_y \otimes \hat{\sigma}_x, \\
\hat{S}_4 &= -\hat{\sigma}_y \otimes \hat{\sigma}_0 \otimes \hat{\sigma}_y, & \hat{S}_5 &= -\hat{\sigma}_y \otimes \hat{\sigma}_z \otimes \hat{\sigma}_0, & \hat{S}_6 &= -\hat{\sigma}_x \otimes \hat{\sigma}_x \otimes \hat{\sigma}_x, & \hat{S}_7 &= +\hat{\sigma}_x \otimes \hat{\sigma}_y \otimes \hat{\sigma}_z.
\end{aligned} \quad (31)$$

The order of the qubits, upon which the Pauli matrices in Eq. (31) act, is $\text{DE} - 2 - 1$. We shall refer to these operators as "stabilizers".[27]

Importantly, the cluster state $|\psi\rangle_{\text{C}}$ is an eigenstate with eigenvalue $+1$ of each of the $\hat{S}_i$, i.e., $\hat{S}_i |\psi\rangle_{\text{C}} = |\psi\rangle_{\text{C}}$. Since the operators are mutually commuting, there is a common diagonalizing basis $\{|\psi\rangle_i\}_{i=0}^7$. We choose $|\psi_0\rangle = |\psi\rangle_{\text{C}}$. Since $\hat{S}_i^2 = \mathbb{I}$, the eigenvalues of the stabilizers are $\pm 1$. Writing $\hat{S}_i |\psi_j\rangle = \lambda_{ij} |\psi_j\rangle$ with $\lambda_{ij} \in \{-1, +1\}$, the stabilizers $\hat{S}_i$ can be related to the spectral projection operators $\hat{P}_i = |\psi_i\rangle\langle\psi_i|$ via

$$\hat{S}_i = \sum_{j=0}^7 \lambda_{ij} \hat{P}_j, \qquad \lambda = \begin{pmatrix} +1 & +1 & +1 & +1 & +1 & +1 & +1 & +1 \\ +1 & +1 & +1 & +1 & -1 & -1 & -1 & -1 \\ +1 & +1 & -1 & -1 & +1 & +1 & -1 & -1 \\ +1 & +1 & -1 & -1 & -1 & -1 & +1 & +1 \\ +1 & -1 & +1 & -1 & +1 & -1 & +1 & -1 \\ +1 & -1 & +1 & -1 & -1 & +1 & -1 & +1 \\ +1 & -1 & -1 & +1 & +1 & -1 & -1 & +1 \\ +1 & -1 & -1 & +1 & -1 & +1 & +1 & -1 \end{pmatrix}. \quad (32)$$

We denote $s_i = \langle \hat{S}_i \rangle = \text{Tr}(\hat{\rho}^{(\text{DE}-2-1)} \hat{S}_i)$ and $p_j = \text{Tr}(\hat{\rho}^{(\text{DE}-2-1)} \hat{P}_j)$. The fidelity, as in Eq. (29), is then given by $\mathscr{F}(\hat{\rho}^{(\text{DE}-2-1)}, |\psi\rangle\langle\psi|_{\text{C}}) = p_0$. Since the matrix in Eq. (32) is invertible, measuring all the values $s_i$ would suffice to determine the probabilities $p_i$ and in particular the fidelity, given by $p_0$. In practice, as $\hat{S}_6, \hat{S}_7$ were not measured, the probabilities $\{p_i\}$ cannot be deduced directly. It turns out, however, that the 6 known values together with the physical requirement that probabilities are non-negative gives a strong restriction on possible values of $p_0$. Thus, for example, the identity $p_0 - p_3 = \frac{1}{4}(s_2 + s_3 + s_4 + s_5)$ together with the relation $p_3 \geq 0$ gives one non-trivial lower bound on the fidelity $p_0$.

We consider the intersection of the simplex of physical probabilities, $p_i \geq 0$ and $\sum_i p_i = 1$, with the confidence ellipsoid corresponding to our measurements. Specifically, we define the function

$$V[p_0, \ldots, p_7] = \sum_{i=1}^5 \left( \frac{s_i - s_i^{\text{meas}}}{\Delta s_i} \right)^2, \quad (33)$$

where $s_1, \ldots, s_5$ are considered as functions of $p_0, \ldots, p_7$ through Eq. (32), and $s_i^{\text{meas}}, \Delta s_i$ are the measured values and their statistical errors.

The measured expectation values of the stabilizers, $s_i^{\text{meas}}$, $i = 1, \ldots, 5$ were calculated from the reconstructed density matrices in Fig. S8. For example, to measure the stabilizer $\hat{S}_2$ one applies

$$s_2^{\text{meas}} = \text{Tr}[\hat{\rho}\hat{S}_2] = \text{Tr}\left[-(\hat{\rho}_{+Z}^{\text{meas}} - \hat{\rho}_{-Z}^{\text{meas}})(\hat{\sigma}_x \otimes \hat{\sigma}_z)\right].$$

Similar relations hold for $\hat{S}_{1,3,4,5}$; these are given in Table S2.

Using the results in Table S2, and demanding $\sqrt{V} < r$ for various values of $r$, gives $p_0 \in [0.542 - 0.035r, 0.694 + 0.022r]$. To obtain this result, we numerically calculated the maximum and minimum of $p_0$ under the constraint $\sqrt{V} \leq r$ for each value of $r$. In particular, the region $p_0 = \mathscr{F}(\hat{\rho}^{(\text{DE}-2-1)}, |\psi\rangle\langle\psi|_{\text{C}}) \leq \frac{1}{2}$ in which there is no genuine 3-qubit entanglement is 1.215 standard deviations away from the measured results.

### 8.2 Entanglement Witnesses

We can also demonstrate tripartite entanglement using an entanglement witness. In close analogy to the case of two qubit states, we define an entanglement witness[19] $\hat{W}$ for a genuine 3-qubit entangled state as a Hermitian quantum operator such that and $\text{Tr}[\hat{W}\hat{\zeta}] \geq 0$ for any bi-separable state $\hat{\zeta}$ (see opening paragraph of Sec. 8 for definition), while $\text{Tr}[\hat{W}\hat{\rho}] < 0$ for some entangled state $\hat{\rho}$.

Using the methods presented in Ref.[27] of constructing witness operators using the stabilizer formalism, we define

$$\begin{aligned}\hat{W} &= 2\mathbb{I} - \hat{S}_5(\mathbb{I} + \hat{S}_6)(\mathbb{I} + \hat{S}_7) \\ &= 2\mathbb{I} - \hat{S}_2 - \hat{S}_3 - \hat{S}_4 - \hat{S}_5,\end{aligned} \quad (34)$$

where the stabilizers $\hat{S}_j$ where defined in Eq. (31). The operator $\hat{W}$ is a "Mermin-type" witness, as in Eq. (28) of Ref.[27], but with a different choice of stabilizers. As such, $\text{Tr}[\hat{W}\hat{\zeta}] \geq 0$ for any bi-separable state, and therefore detects genuine three qubit entanglement.[27]

Using the measured values of the stabilizers, as given in Table S2, we obtain expectation value $\langle \hat{W} \rangle = -0.170 \pm 0.138$, which is 1.23 standard deviations above the separability threshold, thus proving a genuine 3 qubit entanglement.

## 9 Fidelity

The fidelity $\mathscr{F}$ between two Hermitian positive matrices $A$ and $B$ is defined as[28]

$$\mathscr{F}(A,B) = \text{Tr}\left[\sqrt{\sqrt{B}A\sqrt{B}}\right]^2 \Big/ \Big(\text{Tr}[A]\,\text{Tr}[B]\Big). \quad (35)$$

The fidelity between two quantum states $\hat{\rho}_1$ and $\hat{\rho}_2$ is given by $\mathscr{F}(\hat{\rho}_1,\hat{\rho}_2)$. The fidelity between two process maps $\Phi_1$ and $\Phi_2$ is given by $\mathscr{F}(C^{\Phi_1}, C^{\Phi_2})$, where $C^\Phi$ is the Choi matrix corresponding to $\Phi$, see Eq. (5).

### 9.1 Error Calculation

For estimating the error in the calculated fidelity between density matrices, we followed the procedure and notations outlined in Ref.[21]. For any function $g(\hat{\rho})$ of the density matrix, its error is given by $(\Delta g)^2 = \sum_\nu \left(\frac{\partial g}{\partial s_\nu}\right)^2 \Lambda_\nu$, where $s_\nu = \langle \psi_\nu | \hat{\rho} | \psi_\nu \rangle$ is the normalized number of measured coincidences for a particular measurement projection $|\psi_\nu\rangle$, and $\Lambda_\nu$ accounts for the errors in $s_\nu$ resulting from the photon statistics and uncertainty in the polarization projections. The measured density matrix $\hat{\rho}$ is also defined by $\hat{\rho} = \sum_\nu M_\nu s_\nu$, where $M_\nu$ are the auxiliary matrices as defined in Ref.[21].

For calculating the fidelity of $\hat{\rho}$ to a given density matrix $\hat{\tau}$, explicitly $\mathscr{F}(\hat{\rho},\hat{\tau}) = \text{Tr}\left[(\sqrt{\hat{\tau}}\hat{\rho}\sqrt{\hat{\tau}})^{1/2}\right]^2 \doteq \text{Tr}[\sqrt{\hat{\omega}}]^2$, we define auxiliary $\mathscr{F}' = \sqrt{\mathscr{F}}$, whose derivative is given by $\frac{\partial \mathscr{F}'}{\partial s_\nu} = \text{Tr}\left[\frac{1}{2}(\sqrt{\hat{\omega}})^{-1}\sqrt{\hat{\tau}}M_\nu\sqrt{\hat{\tau}}\right]$, where $(\sqrt{\hat{\omega}})^{-1}$ is the pseudo-inverse of $\sqrt{\hat{\omega}}$. The total error is given by $\Delta \mathscr{F} = 2\mathscr{F}'\Delta\mathscr{F}'$.


\setcounter{page}{\value{mainpage}}    

\end{document}